\documentclass{aa}

\usepackage[dvips]{graphics}

\usepackage{natbib,rotate,lscape,epsf,psfig,graphicx,amsmath,amssymb,verbatim}
\usepackage{natbib,rotate,lscape,epsf,graphicx,amsmath,amssymb,verbatim}
\usepackage{latexsym,longtable,colortbl,lscape}
\usepackage{aalongtable}

\definecolor{gris}{gray}{0.8}
\usepackage{txfonts}
\begin{document}

\title{A survey of submillimeter C and CO lines in nearby galaxies.}
\subtitle{submitted}

\author{E.Bayet\inst{1}, M.Gerin\inst{1}, T.G.Phillips\inst{2}
\and A. Contursi\inst{3} }

\offprints{E.Bayet, \email{estelle.bayet@lra.ens.fr}}

\institute{Laboratoire de Radioastronomie (LRA), Observatoire de
Paris and Ecole Normale Sup\'erieure, 24 rue Lhomond, F-75005 Paris,
France (CNRS-UMR 8112) \and California Institute of Technology, Downs
Laboratory of Physics 320-47, Pasadena, CA 91125, USA \and Max Planck
Institute fuer Extraterrestrische Physik, Giessenbachstrasse, 85748
Garching, Germany}

\date{Received ../../06;}
\abstract{}{}{}{}{}
   \abstract
   {}
   {While the search for molecular gas in distant galaxies is based on
   the detection of submillimeter CO rotational lines, the current CO
   surveys of nearby galaxies are restricted to the millimeter CO
   lines. The submillimeter CO lines are formed in warm and dense
   molecular gas and are therefore sensitive to the physical
   conditions whereas the CO
   ($J=1\rightarrow 0$) line is a tracer of the total molecular gas
   mass. In order to be able to compare the properties of molecular
   gas in nearby and distant galaxies, we have observed C and CO
   submillimeter lines (including the $^{12}$CO(6-5) and
   $^{12}$CO(7-6) lines) in a sample of nearby galaxies using the
   \textit{Caltech Submillimeter Observatory (CSO)}.}
   {We have obtained a complete view of the CO
   cooling curve (also called CO spectral energy distribution)
   by combining the submillimeter CSO data with previous observations
   found in the literature. We
   made use of Large Velocity Gradient (LVG) models to analyse the
   observed CO cooling curve, predict CO line intensities from $J=1
   \rightarrow 0$ to  $J=15 \rightarrow 14$ in the studied galaxies,
   and derive the physical properties of the warm and dense  molecular
   gas : the kinetic temperature (T$_{K}$); the gas density
   (n(H$_{2}$)); the CO column density divided by the line width
   N($^{12}$CO)/$\Delta v$. The predictions for the line intensities
   and for the total CO cooling power, obtained from LVG modelling
   have been compared with predictions from Photo Dissociation Regions
   (PDR) models.}
   {We show how the CO SED varies according to the galaxy
   star forming activity. For active nuclei, the peak is located near
   the $^{12}$CO(6-5) or $^{12}$CO(7-6)
   rotational lines, while, for normal nuclei, most of the energy is
   carried by the $^{12}$CO(4-3) and $^{12}$CO(5-4) lines. Whatever the
   spectral type of the nucleus, the observed C cooling rate is lower
   than the observed CO cooling rate (by a factor of $\geqslant$ 4). The CO
   cooling curve of nearby starburst galaxies (e.g. NGC~253) has a
   quite similar shape to the CO cooling curve of distant
   galaxies. Therefore, the CO cooling curves are useful
   diagnostics for the star forming activity in distant objects. }
   {}

\keywords{Galaxies: starburst-ISM-nuclei -- Galaxies: individual: The
Antennae-Arp 220-Centaurus A-Cloverleaf QSO-IC 10-IC 342-IRAS
10565+2448-Milky Way-M 51-M 82-M 83-Markarian 231-NGC 3079-NGC
4736-NGC 6090-NGC 6946 -- Submillimeter -- ISM: molecules}

\titlerunning{Submillimeter C and CO lines survey}
\authorrunning{E.Bayet et al.}


\section{Introduction}\label{sec:intro}

It is well known that the far infrared fine-structure lines of
abundant elements (Oxygen, Carbon, Nitrogen, Silicon, Sulfur, etc),
either in their neutral or ionized states, contribute most of the gas
cooling of the interstellar medium in galaxies \citep{Holl99,
Goic05}. Far infrared fine-structure lines from ionized atoms are
useful tracers of HII regions (e.g. \citealt{Goic05}). For the
neutral ISM, the main cooling lines are those of ionized carbon [CII]
and neutral oxygen [OI]. By analyzing a set of ISO observations of
external galaxies, \citet{Malh01} have concluded that PDRs contribute
to a large fraction of the emission of the [CII] and [OI] lines. In
molecular gas, the cooling radiation due to atomic carbon and carbon
monoxide is significant. The theoretical predictions \citep{Gold78,
Holl99} have been confirmed by the COBE-FIRAS and ISO-LWS
observations of the Milky Way : apart from [CII], [OI], [OII] and
[NII], the most intense lines are from C and CO. The relative
contributions of the different lines of C and CO vary along the
Galactic plane : the brightest lines and more excited states are seen
towards the Galactic Center, while the rest of the disk shows lower
excitation. Also, C contributes less in proportion to the total
cooling towards the Galactic Center than towards the disk
\citep{Benn94, Fixs99}.

The CO cooling is typically provided by the submillimeter lines, with
rotational quantum numbers between 3 and 8 \citep{Baye04}. Such
observations can be performed from high altitude dry mountain sites,
such as the Mauna Kea summit in Hawaii. Previous studies using the
\textit{James Clerk Maxwell Telescope} (\textit{JCMT}) and the
\textit{CSO} have shown that the $^{12}$CO(4-3) line is generally
bright in galaxy nuclei \citep[...]{Gues93, Isra95, Isra01, Isra03,
Kram05}. Much less information is available on the other high
frequency CO lines. The first report of extragalactic $^{12}$CO(6-5)
detections was published by \citet{Harr91} 15 years ago. In the past
years new analyses of the $^{12}$CO(6-5) and $^{12}$CO(7-6) emission
have been presented by \citet{Ward01, Ward03} for M 82, and by
\citet{Brad03} and \citet{Baye04} for NGC 253 and Henize 2-10. A
first comparison of the CO line spectrum, also called "CO SED" (see
\citealt[...]{Weis05, Mao00}) in star forming galaxies is presented
by \citet{Baye04}. They show that the CO line spectrum is very
similar in the two star forming galaxy nuclei, NGC 253 and Henize
2-10. Distant starburst galaxies seem also to share the same CO
spectrum \citep[...]{Cox02, Bert03, Cari04, Pety04, Walt04, Cari05}.

Although contributing less than CO, atomic carbon is an important
coolant in the ISM \citep{Geri00}. While observations of the ground
state line ($^3$P$_1$ - $^3$P$_0$ at 492 GHz) can be found in the literature
for a few tens of sources (e.g. \citealt{Geri00, Isra02}) there are very
few reported detections of the excited line ($^3$P$_2$ - $^3$P$_1$ at 809
GHz) although the contribution of the latter line to the neutral
carbon cooling is at least similar to the contribution of the ground
state line \citep{Stut97, Baye04}.

As potential tracers of the gas cooling, submillimeter C and CO lines
are expected to provide information on the gas heating rate, which is
dominated by the incident FUV radiation, mainly due to massive and
young stars. Therefore, the molecular cooling lines are expected to
provide information on the galaxy star forming activity, as do the
fine structure lines in the far infrared. In order to have a full
picture of the CO cooling, the contribution of missing CO lines
(blocked by the Earth's atmosphere) can be predicted from the series
of observed lines, using state-of-the-art radiative transfer models.
The method has been presented in our previous paper \citep{Baye04}.
In this paper, we present in this paper results of a survey of the C
and CO submillimeter lines in a sample of nearby galaxies. The data
are used for the two following purposes : {\em i) determination of
the molecular cooling rate in galaxies of different morphological
type; ii) study the shape of the CO cooling curves obtained in the
target galaxies}. From this analysis, we show that the combined
information on C and CO submillimeter line spectra can be used as a
powerful diagnostic of galaxy star forming activity.

The galaxy sample is presented in section 2, the observations
parameters are described in section 3, while the resulting spectra
and maps are introduced in section 4. We discuss in section 5 how we
use LVG and PDR models for fitting the series of observed CO lines.
In section 6, we compare results obtained for the center of Milky Way
and for the Cloverleaf QSO with those derived from this work for our
galaxy sample. The main conclusions are summarized in section 7.


\section{The sample}\label{sec:sample}

We have selected galaxies which are bright in the $^{12}$CO(1-0) and
$^{12}$CO(2-1) lines and are nearby (distance less than 20 Mpc,
except for the two ULIRGs Arp220 and Markarian 231). Galaxies have
also been selected to have a large variety of galaxy types. The
sample includes normal spiral galaxies (IC 342, M 51, NGC 4736, NGC
6946), starburst galaxies (M 82, M 83, NGC 253, NGC 3079), irregular,
star forming galaxies (IC 10, Henize 2-10), interacting galaxies (The
Antennae, NGC 6090), ULIRGs (Arp 220, IRAS 10565+2448, Markarian 231)
and the elliptical galaxy Centaurus A. In this work, we analyzed two
positions in the Antennae galaxies : the nucleus of the northern
component, NGC 4038, and a position named ``Overlap'' hereafter,
which corresponds to the position of the most massive H$_{2}$
concentration, not very far from the nucleus of NGC 4039 (The
coordinates of NGC 4038 and Overlap are listed in Table
~\ref{tab:prop}). Intense MIR emission due to star formation has been
detected at the Overlap position \citep{Vigr96} as well as bright CO
lines (GMC4-5 in \citealt{Wils00}) .

Properties of the sample galaxies are summarized in Table
~\ref{tab:prop}. Although it is not a complete sample (because these
observations are difficult and time consuming) it includes
representative types of nearby galaxies.

\begin{table*}
    \caption{Basic properties of the sample galaxies.}\label{tab:prop}
    \begin{center}
        \begin{tabular}{l c c c c c c c}\hline\hline
        & Type & RA(1950) & DEC(1950) & Dist.& Velocity &
        Optical &  Metallicity:  \\
        &&&&&LSR&size$^{a}$&12+log$\frac{O}{H}$\\
        & & & & (Mpc) & (kms$^{-1}$) & & \\
        \hline
        IC 10 & dIrr IV/BCD$^{a}$ & 00:17:44.0 & 59:00:18.0 & 1$^{1}$
        & -344 & 6.8'$\times$5.8' & 8.31$\pm$0.20$^{14}$ \\
        \hline
        NGC 253$^{b}$ & SAB(s)c;H$_{\text{II}}$, Sbrst & 00:45:05.7 &
        -25:33:38.0 & 2.5$^{18}$ & 240 & 27.5' $\times$ 6.8' &
        8.99$\pm$ 0.31$^{16}$\\
        \hline
        IC 342 & SAB(rs)cd HII$^{a}$ & 03:41:57.2 & 67:56:27.0 &
        1.8$^{2}$ & 35 & 21.4'$\times$20.9'& $\approx$9.30$^{15}$\\
        \hline
        Henize 2-10$^{b}$ & I0 pec, Sbrst & 08:34:07.2 & -14:26:06.0 &
        6$^{17}$ & 850 & 30'' $\times$ 40 '' & $\approx$8.93$^{16}$ \\
        \hline
        M 82 & I0;Sbrst HII$^{a}$ & 09:51:43.8 & 69:55:00.9 &
        3.2$^{3}$ & 200 & 11.2'$\times$4.3' & 9.00$\pm$0.12$^{14}$ \\
        \hline
        NGC 3079 & SB(s)c;LINER Sy2$^{a}$ & 09:58:35.0 & 55:55:15.4
        &15.6$^{4}$ & 1331 & 7.9'$\times$1.4' & - \\
        \hline
        IRAS 10565+2448& LINER HII$^{a}$ & 10:56:36.2 & 24:48:40.0 & 172$^{5}$
        & 12923 &0.4'$\times$0.3' & - \\
        \hline
        NGC 4038 & SB(s)m pec$^{a}$ & 11:59:19.0 & -18:35:23.0 &
        13.8$^{6}$ & 1634 & 5.9'$\times$3.2' & - \\
        \hline
        Overlap & SA(s)m pec$^{a}$ & 11:59:21.1 & -18:36:17.0 &
        13.8$^{6}$ & 1510 & 3.1'$\times$1.6' & - \\
        \hline
        NGC 4736 & (R)SA(r)ab;Sy2 LINER$^{a}$ & 12:48:32.4 &
        41:23:28.0 & 4.3$^{7}$ & 314 & 11.2'$\times$9.1'&
        9.01$\pm$0.17$^{16}$ \\
        \hline
        Mrk 231 & SA(rs)c? pec Sy1$^{a}$ & 12:54:05.0 & 57:08:39.0 &
        173.9$^{8}$ & 12650& 1.3'$\times$1.0' & -\\
        \hline
        Centaurus A & S0 pec Sy2$^{a}$ & 13:22:31.6 & -42:45:32.0 &
        3.5$^{9}$ & 550 & 25.7'$\times$20.0'& - \\
        \hline
        M 51 & SA(s)bc pec; HII Sy2.5$^{a}$& 13:27:46.1 & 47:27:14.0
        & 9.6$^{10}$ & 470 & 11.2'$\times$6.9'& 9.23$\pm$0.12$^{16}$ \\
        \hline
        M 83 & SAB(s)c; HII Sbrst$^{a}$ & 13:34:11.3 & -29:36:42.6 &
        3.5$^{11}$ & 516 & 12.9'$\times$11.5' & 9.16$\pm$0.12$^{16}$  \\
        \hline
        Arp 220 & S?;LINER; HII Sy2$^{a}$ & 15:32:46.7 & 23:40:08.0 &
        77$^{10}$ & 5450 & 1.5'$\times$1.2' & - \\
        \hline
        NGC 6090 & Sd pec HII$^{a}$ & 16:10:23.9 & 52:35:11.0 &
        118$^{12}$  & 8831 & 2.8'$\times$1.5' & - \\
        \hline
        NGC 6946 & SAB(rs)cd HII$^{a}$ & 20:33:48.8 & 59:58:50.0 &
        5.5$^{13}$  & 50 & 11.5'$\times$9.8' & 9.06$\pm$0.17$^{16}$ \\
        \hline
        \end{tabular}
    \end{center}
    \begin{minipage}{18cm}
    References: 1: Adopted value (see text); 2: \citet{McCa89, Kara93};
    3: \citet{Dumk01} consistent with the value of \citet{Tamm68}; 4:
    \citet{Sofu99}; 5: \citet{Glen01}; 6: \citet{Savi04}; 7:
    \citet{Tull87}; 8: \citet{Brya99}; 9: \citet{DeVa79}; 10:
    \citet{DeVa91}; 11: \citet{Thim03}; 12: Redshift from
    \citet{Geri99} with H$_{0}$ = 75 kms$^{-1}$Mpc$^{-1}$; 13:
    \citet{Tull88}; 14: \citet{Arim96}; 15: \citet{Vila92, Garn98b};
    16: \citet{Zari94}; 17: \citet{Joha87}; 18 : As in \citet{Maue96}
    (see \citet{Baye04}); $^{a}$: Data from the NED database; $^{b}$:
    See \citet{Baye04} to obtain more information on the properties of
    NGC 253 and Henize 2-10.
    \end{minipage}
\end{table*}


\section{Observations}\label{sec:obs}

\begin{figure}
  \begin{center}
    \epsfxsize=9cm
    \epsfbox{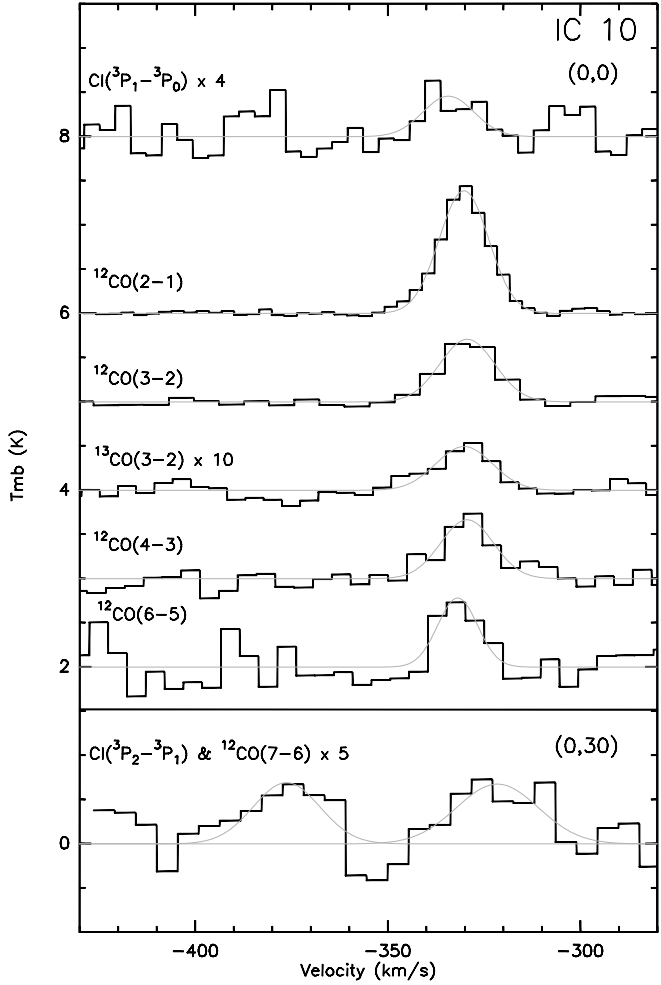}
    \caption{Observed spectra towards IC 10 nucleus
    (see Table ~\ref{tab:obs1}) except for the
    [CI]($^{3}$P$_{2}$-$^{3}$P$_{1}$) and the $^{12}$CO(7-6)
    spectra which correspond to the offset position (0",30"). Velocities
    (horizontal axis) are given in kms$^{-1}$ relative to the
    LSR (V$_{LSR}$) and the line intensities (vertical axis)
    are in units of T$_{mb}$ (K). The grey curves are Gaussian
    fits. The observed line is written above each spectrum.
    In this figure, we present the [CI]($^{3}$P$_{2}$-$^{3}$P$_{1}$)
    and the $^{12}$CO(7-6) lines on
    the same spectrum since they have been observed
    simultaneously. The [CI]($^{3}$P$_{1}$-$^{3}$P$_{0}$)
    spectrum is from \citet{Geri00} but it has been
    analyzed again to obtain an homogeneous dataset
    of observations.}\label{fig:spec_ic10}
  \end{center}
\end{figure}

\begin{figure}
  \begin{center}
    \epsfxsize=9cm
    \epsfbox{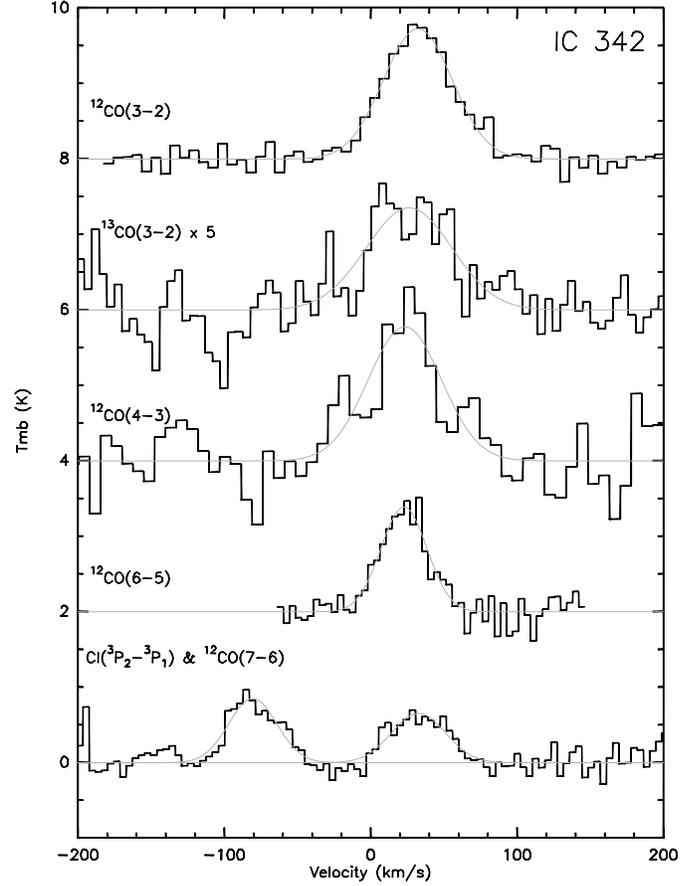}
    \caption{Observed spectra towards IC 342 nucleus (see
    Table ~\ref{tab:obs1}). See the
    caption of Fig. ~\ref{fig:spec_ic10}. We present
    the [CI]($^{3}$P$_{2}$-$^{3}$P$_{1}$) and the
    $^{12}$CO(7-6) lines on the same spectrum since
    they have been observed simultaneously.}\label{fig:spec_ic342}
  \end{center}
\end{figure}

\begin{figure}
  \begin{center}
    \epsfxsize=7cm
    \epsfbox{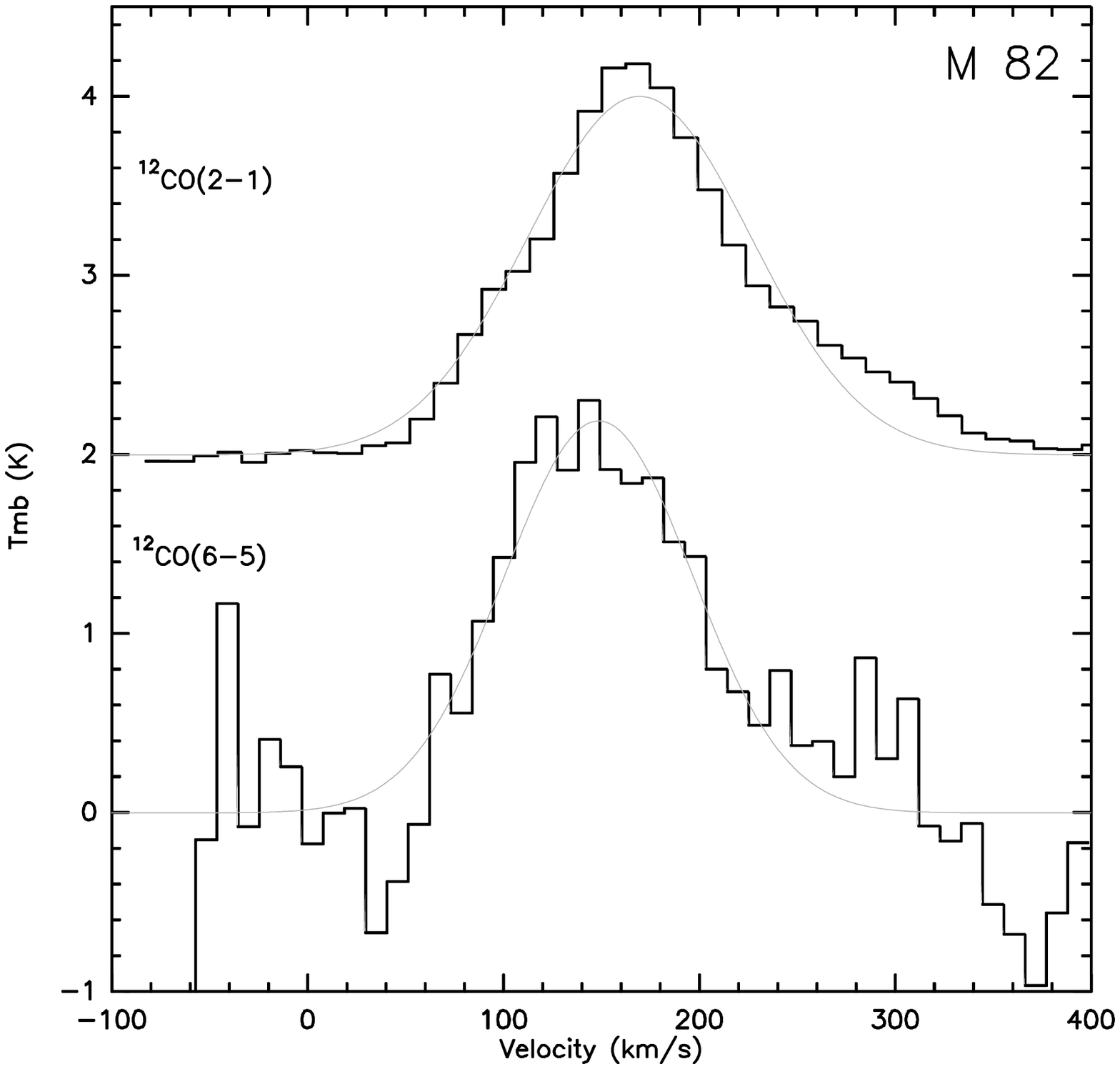}
    \caption{Observed spectra towards M 82 nucleus (see
    Table ~\ref{tab:obs1}). See the
    caption of Fig. ~\ref{fig:spec_ic10}.}\label{fig:spec_m82}
  \end{center}
\end{figure}

\begin{figure}
  \begin{center}
    \epsfxsize=7cm
    \epsfbox{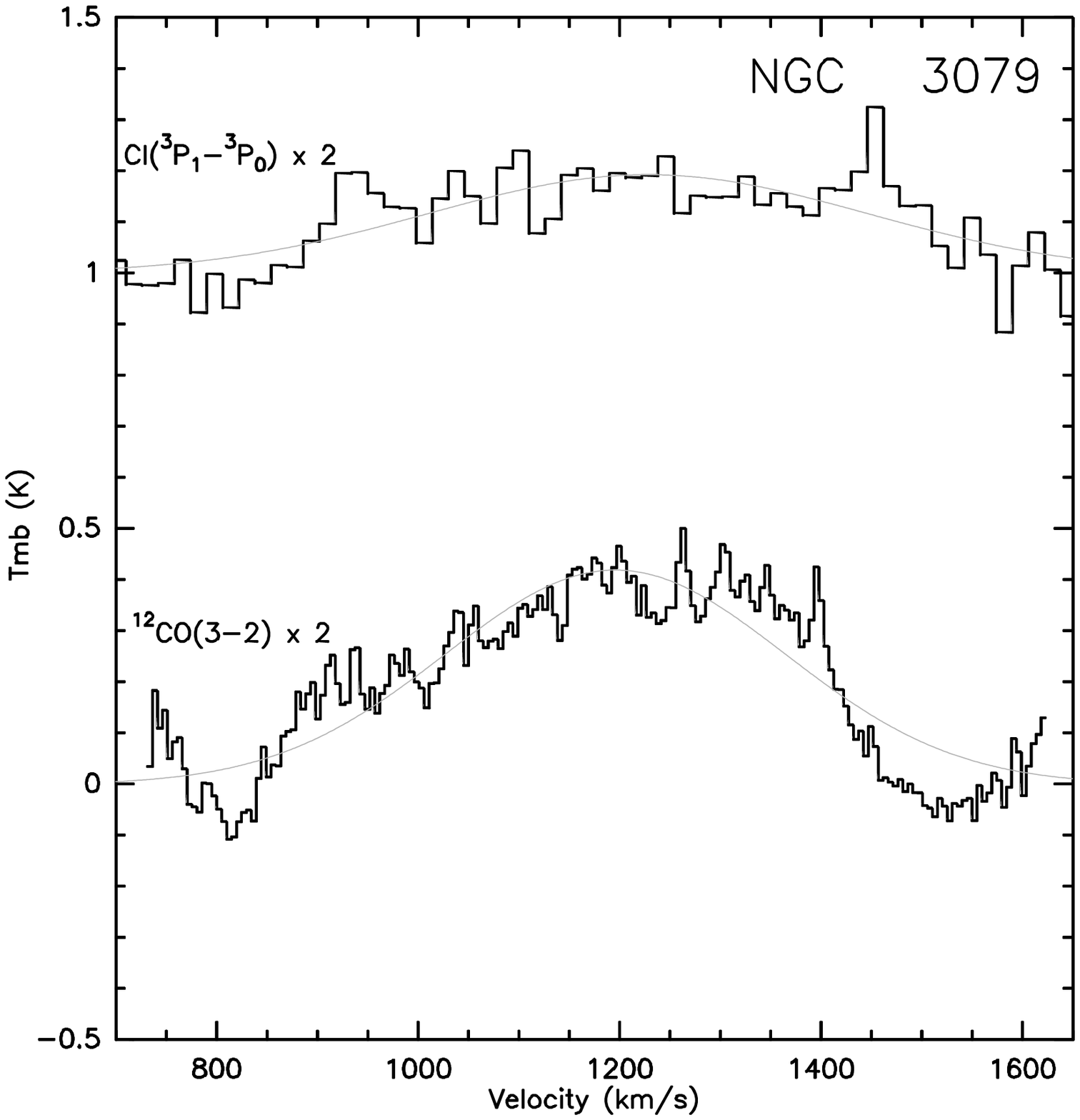}
    \caption{Observed spectra towards NGC 3079 nucleus (see
    Table ~\ref{tab:obs1}). See the
    caption of Fig. ~\ref{fig:spec_ic10}.}\label{fig:spec_n3079}
  \end{center}
\end{figure}

\begin{figure}
  \begin{center}
    \epsfxsize=7cm
    \epsfbox{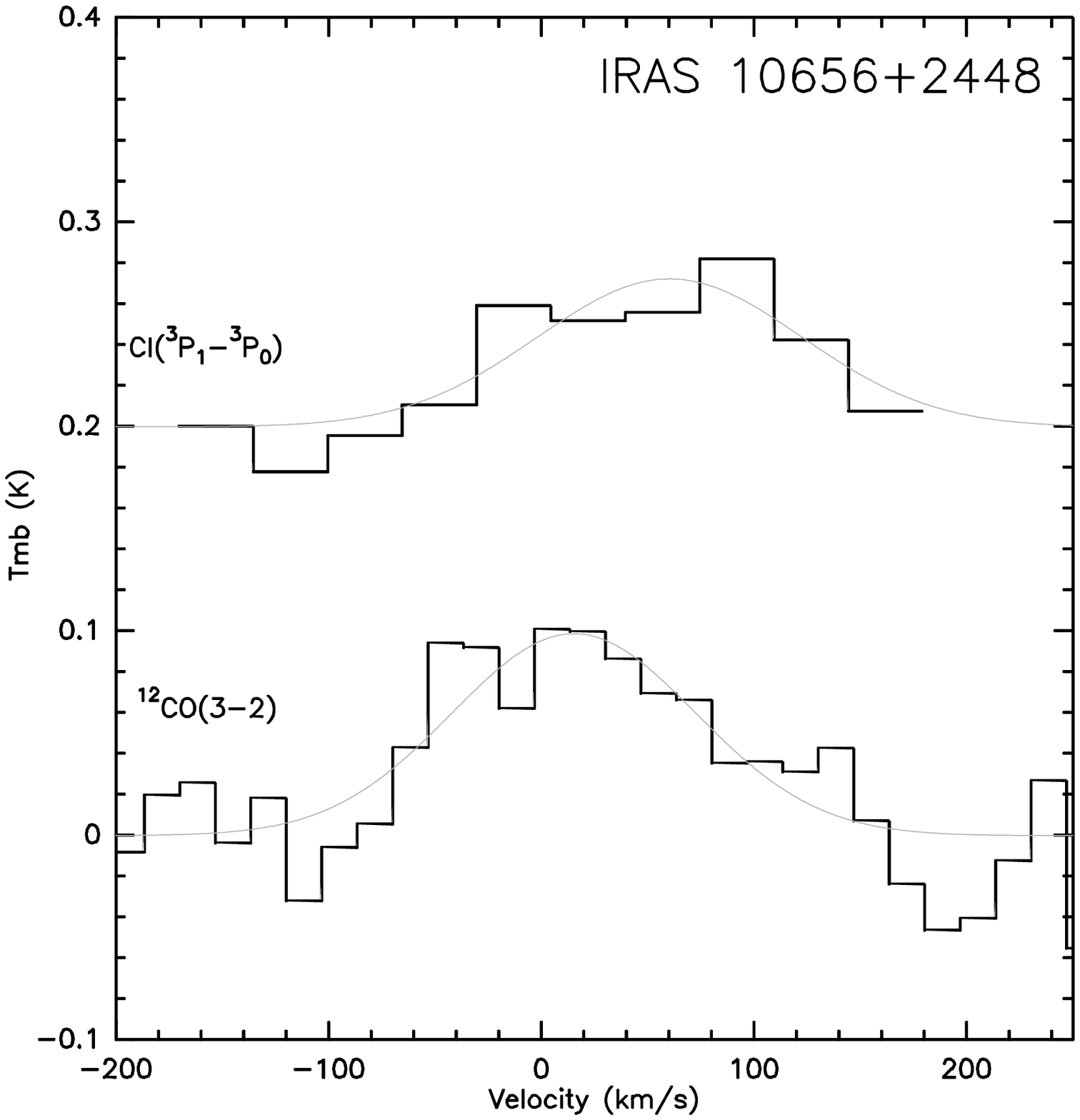}
    \caption{Observed spectra towards IRAS 10565+2448 nucleus (see
    Table ~\ref{tab:obs1}). See the
    caption of Fig. ~\ref{fig:spec_ic10}.}\label{fig:spec_i10656}
  \end{center}
\end{figure}

\begin{figure}
  \begin{center}
    \epsfxsize=9cm
    \epsfbox{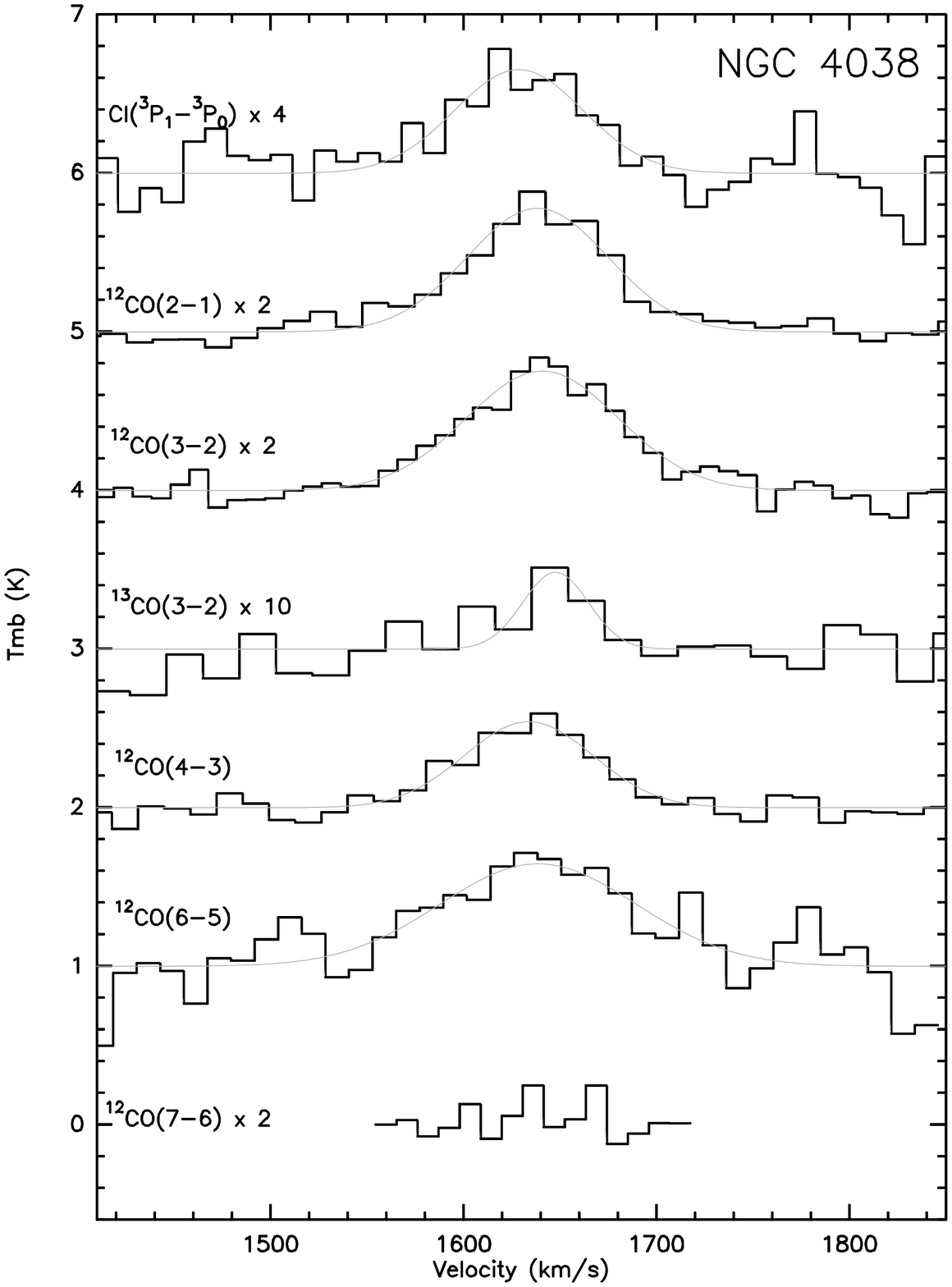}
    \caption{Observed spectra towards NGC 4038 nucleus (see
    Table ~\ref{tab:obs1}). See the
    caption of Fig. ~\ref{fig:spec_ic10}.}\label{fig:spec_ant2}
  \end{center}
\end{figure}

\begin{figure}
  \begin{center}
    \epsfxsize=9cm
    \epsfbox{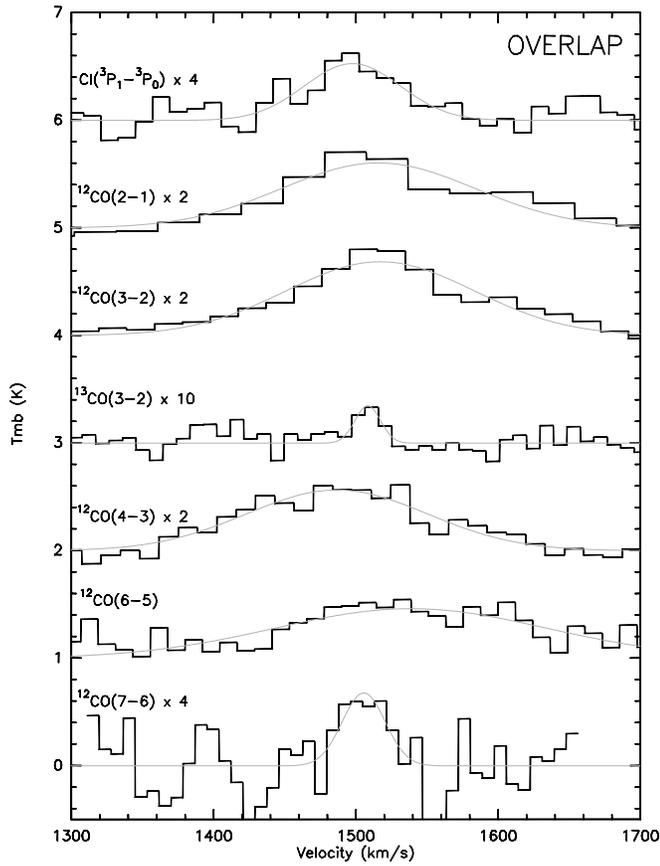}
    \caption{Observed spectra towards the Overlap region (see
    Table ~\ref{tab:obs1}). See the
    caption of Fig. ~\ref{fig:spec_ic10}.}\label{fig:spec_ant1}
  \end{center}
\end{figure}

\begin{figure}
  \begin{center}
    \epsfxsize=7cm
    \epsfbox{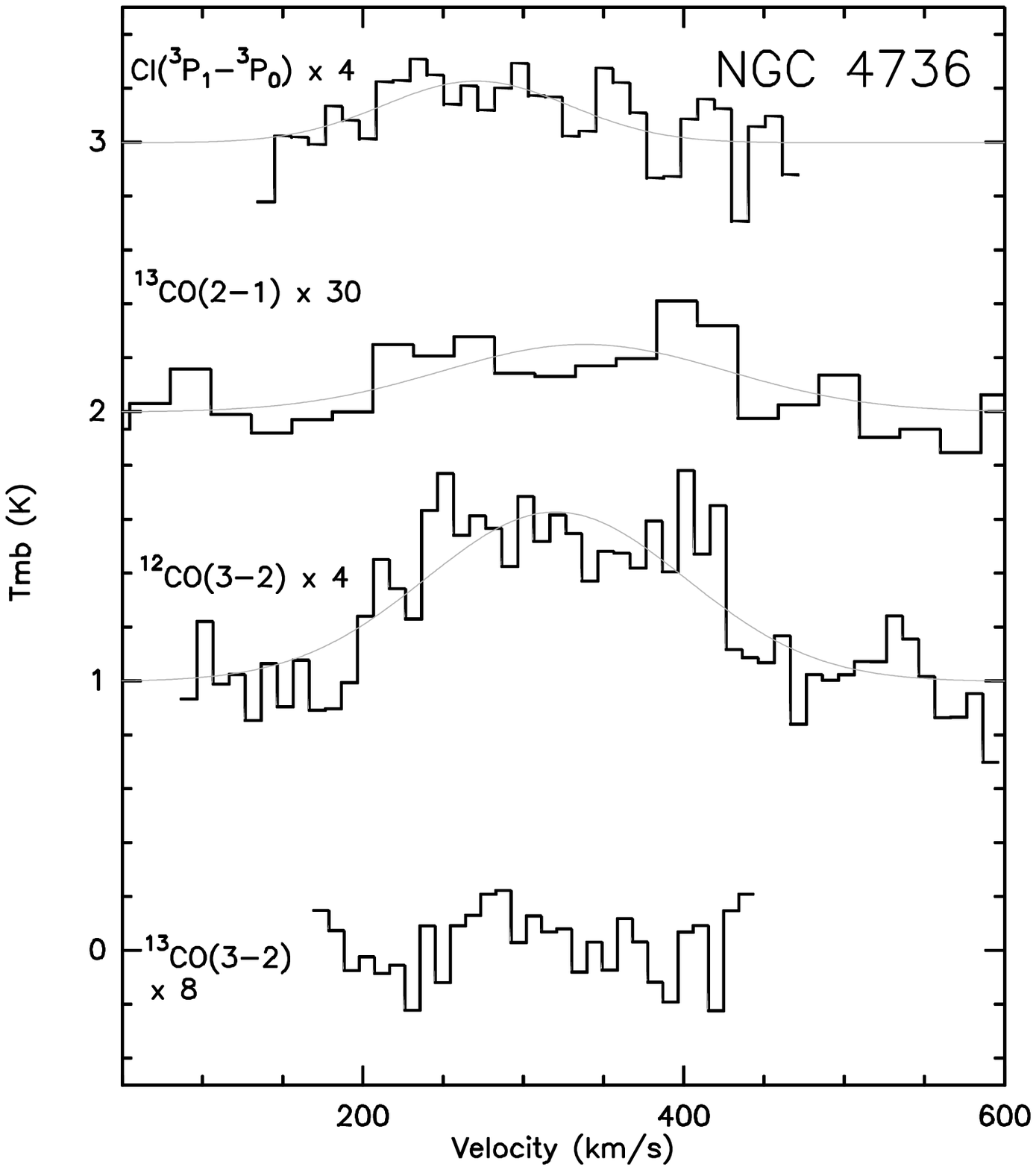}
    \caption{Observed spectra towards NGC 4736 nucleus (see
    Table ~\ref{tab:obs1}). See the
    caption of Fig. ~\ref{fig:spec_ic10}.}\label{fig:spec_n4736}
 \end{center}
\end{figure}

\begin{figure}
  \begin{center}
    \epsfxsize=7cm
    \epsfbox{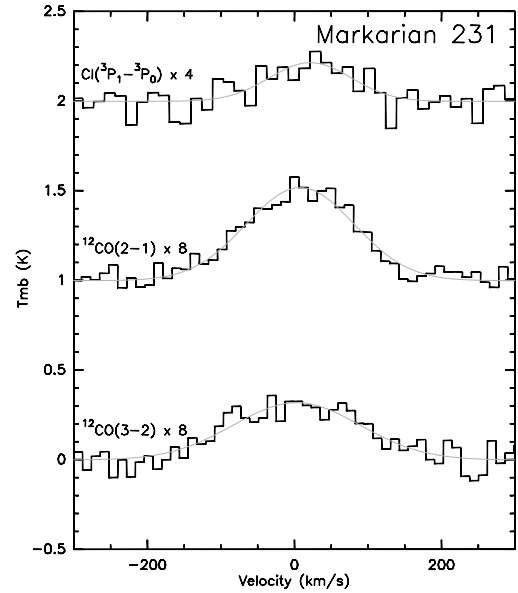}
    \caption{Observed spectra towards Markarian 231 nucleus (see
    Table ~\ref{tab:obs1}). See the
    caption of Fig. ~\ref{fig:spec_ic10}.}\label{fig:spec_mrk231}
  \end{center}
\end{figure}

\begin{figure}
  \begin{center}
    \epsfxsize=7cm
    \epsfbox{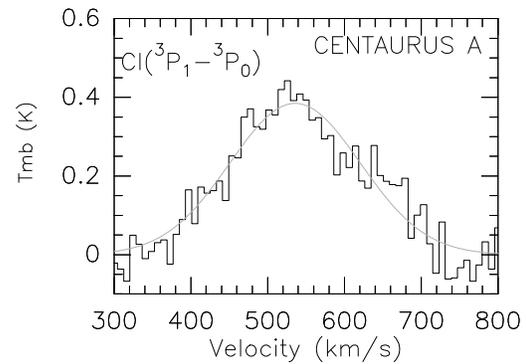}
    \caption{Observed spectra towards Centaurus A nucleus (see
    Table ~\ref{tab:obs1}). See the
    caption of Fig. ~\ref{fig:spec_ic10}.}\label{fig:spec_cena}
  \end{center}
\end{figure}

\begin{figure}
  \begin{center}
    \epsfxsize=9cm
    \epsfbox{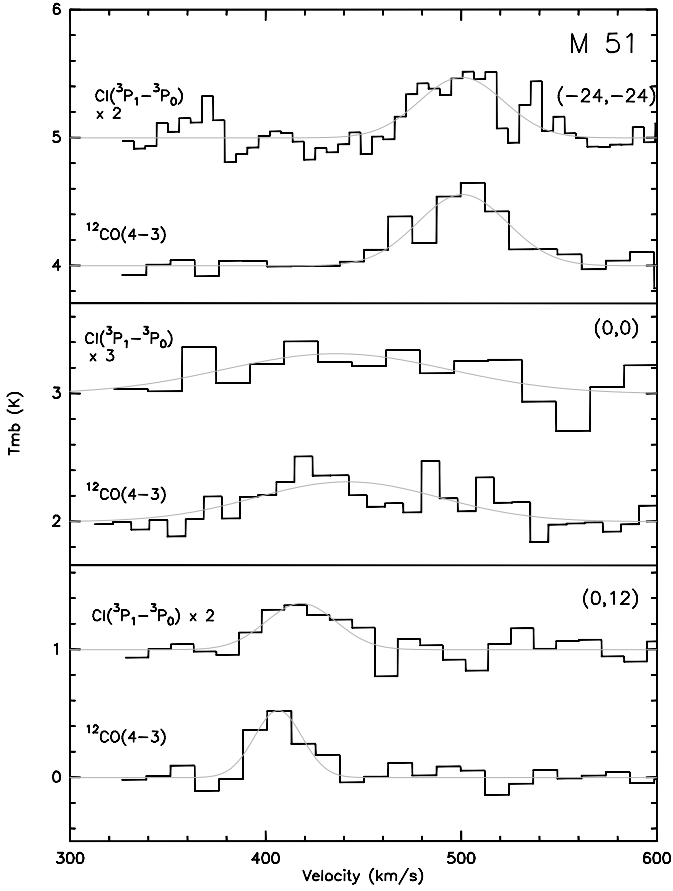}
    \caption{Spectra of M 51 at various offset positions
    relative to the central position (0'',0'') given in
    Table ~\ref{tab:prop} (see also
    Table ~\ref{tab:obs1}) : (-24",-24") (top), (0",0")
    (middle) and (0",12") (bottom). See the caption of
    Fig. ~\ref{fig:spec_ic10}. The [CI]($^{3}$P$_{1}$-$^{3}$P$_{0}$)
    spectrum at the central position (0",0") is from
    \citet{Geri00} but it has been analyzed again to
    obtain an homogeneous dataset of observations.}\label{fig:spec_m51}
  \end{center}
\end{figure}

\begin{figure}
  \begin{center}
    \epsfxsize=9cm
    \epsfbox{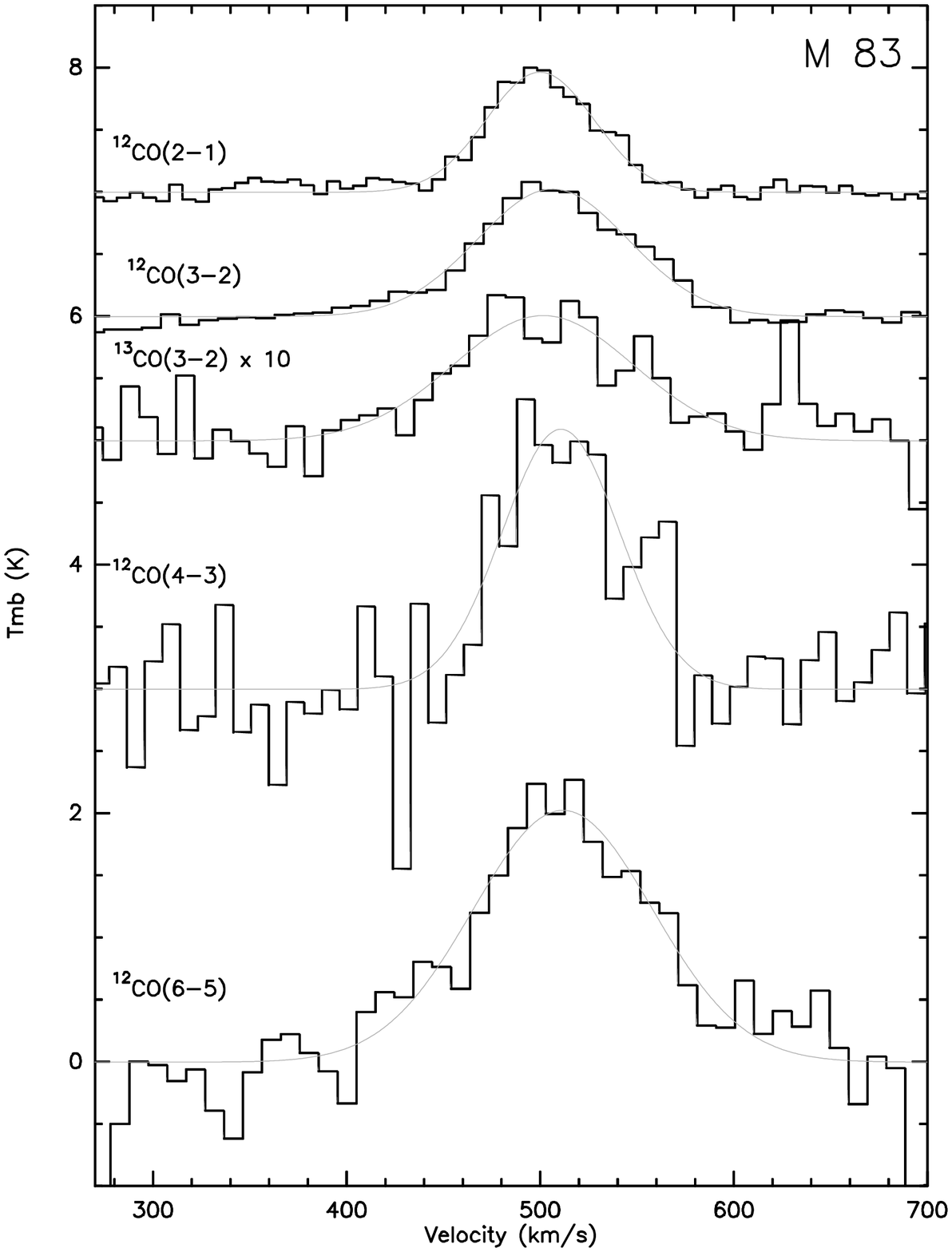}
    \caption{Observed spectra towards M 83 nucleus (see
    Table ~\ref{tab:obs1}). See the
    caption of Fig. ~\ref{fig:spec_ic10}.}\label{fig:spec_m83}
  \end{center}
\end{figure}

\begin{figure}
  \begin{center}
    \epsfxsize=7cm
    \epsfbox{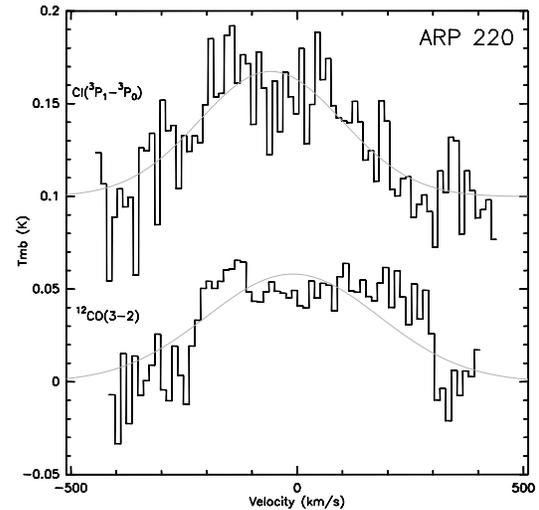}
    \caption{Observed spectra towards Arp 220 nucleus (see
    Table ~\ref{tab:obs1}). See the
    caption of Fig. ~\ref{fig:spec_ic10}. In this figure, both
    spectra are from \cite{Geri98} but they have been analyzed
    again to obtain an homogeneous dataset of
    observations.}\label{fig:spec_arp220}
  \end{center}
\end{figure}

\begin{figure}
  \begin{center}
    \epsfxsize=9cm
    \epsfbox{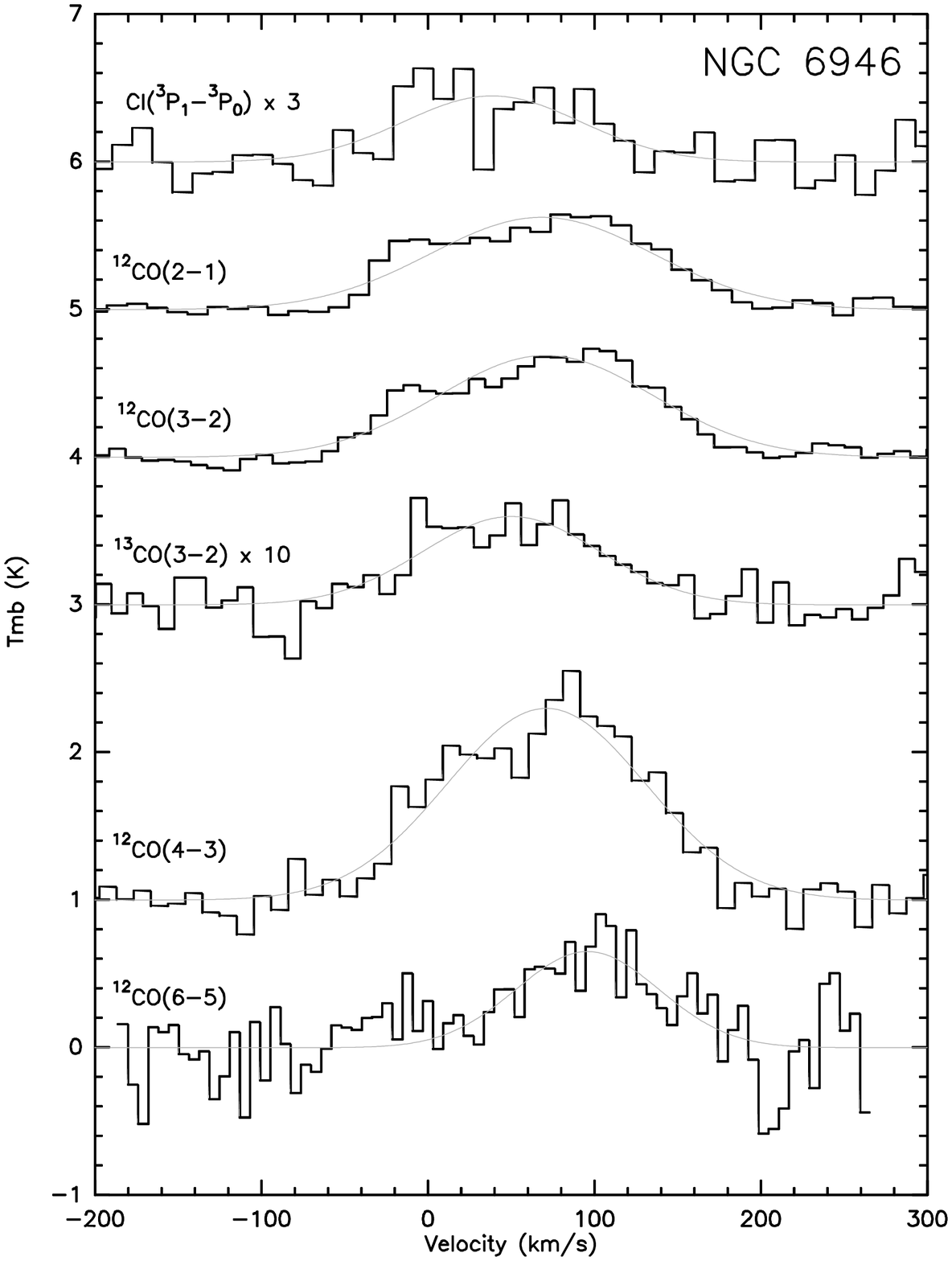}
    \caption{Observed spectra towards NGC 6946 nucleus (see
    Table ~\ref{tab:obs1}). See the
    caption of Fig. ~\ref{fig:spec_ic10}.}\label{fig:spec_n6946}
  \end{center}
\end{figure}

\begin{figure}
  \begin{center}
    \epsfxsize=9cm
    \epsfbox{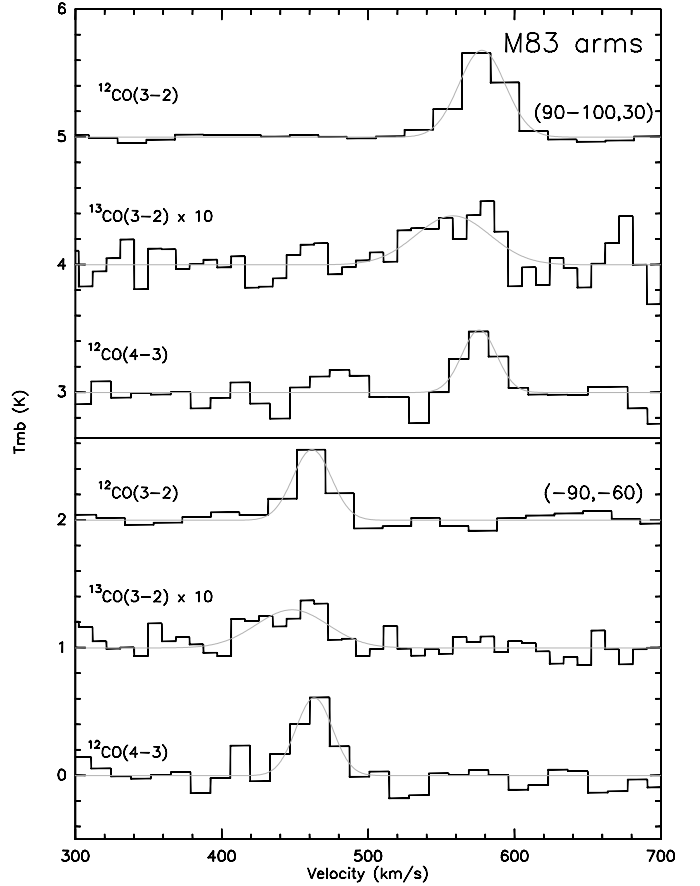}
    \caption{Spectra taken in the M 83 arms at the offset positions
    (90''-100'',30'') (top) and (-90'',-60'') (bottom) (relative to
    the central position (0'',0'') listed in Table ~\ref{tab:prop},
    see also Table ~\ref{tab:obs1}). See the caption
    of Fig. ~\ref{fig:spec_ic10}.}\label{fig:spec_m83_bras}
 \end{center}
\end{figure}

\begin{figure}
  \begin{center}
    \epsfxsize=9cm
    \epsfbox{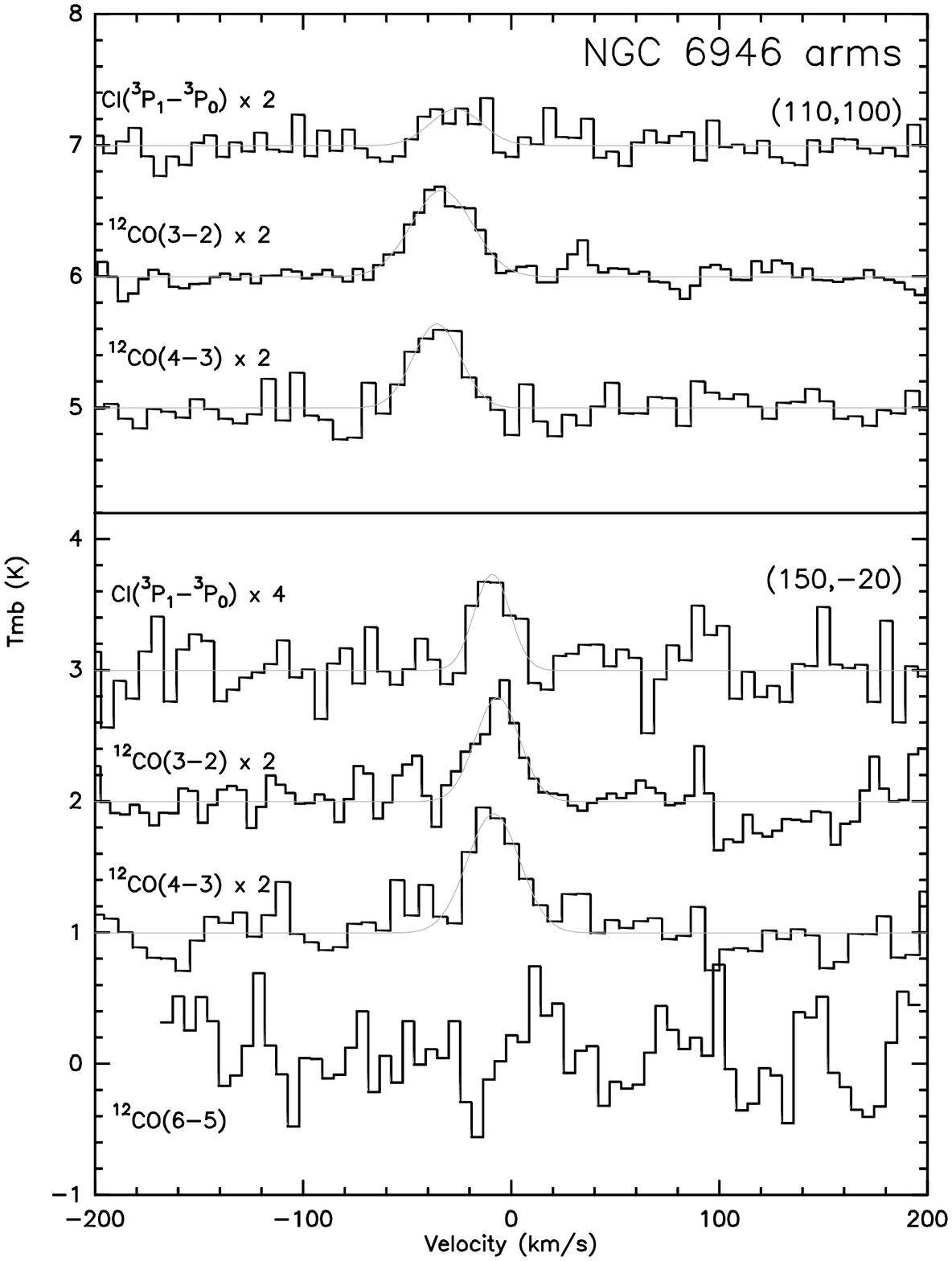}
    \caption{Spectra taken towards the NGC 6946 arms at the
    offset positions (110'',100'') (top) and (150'',-20'') (bottom)
    (relative to the central position (0'',0'') listed in
    Table ~\ref{tab:prop}, see also
    Table ~\ref{tab:obs1}). See the caption of
    Fig. ~\ref{fig:spec_ic10}.}\label{fig:spec_n6946_bras}
  \end{center}
\end{figure}

The observations were made during various sessions at the
\textit{Caltech Submillimeter Observatory} (\textit{CSO}) in Hawaii
(USA) with the Superconducting Tunnel Junction receivers operated in
double-side band mode. The atmospheric conditions varied from good
($\tau_{225} \lesssim 0.1$) to excellent ($\tau_{225}\approx$ 0.06).
We used a chopping secondary mirror with a frequency of around 1 Hz.
We used a 3' chopping throw for the
[CI]($^{3}$P$_{1}$-$^{3}$P$_{0}$), CO(2-1), CO(3-2) and CO(4-3)
lines. There is no sign of contamination by emission in the off
beams. We restricted the chopping throw to 1' for
[CI]($^{3}$P$_{2}$-$^{3}$P$_{1}$), CO(6-5) and CO(7-6) as the
emission is very compact in these lines. Spectra were measured with
two acousto-optic spectrometers (effective bandwidth of 1000 MHz and
500 MHz). The first one has a spectral resolution about 1.5 MHz and
the second one about 2 MHz. The IF frequency of the CSO receivers is
1.5 GHz. The main beam efficiencies ($\eta$) of the CSO were 69.8\%,
74.6\%, 51.5\%, 28\% and 28\% at 230, 345, 460, 691 and 806 GHz
respectively, as measured on planets$^{1}$. For the
[CI]($^{3}$P$_{1}$-$^{3}$P$_{0}$) and the
[CI]($^{3}$P$_{2}$-$^{3}$P$_{1}$) lines, we used receivers at 492 and
809 GHz, so $\eta=$ 51.5\% and 28\%, respectively. We used the ratio
$\frac{1}{\eta}$ to convert T$^{*}_{A}$ into T$_{mb}$. The beam size
at 230, 345, 460, 691 and 806 GHz is 30.5'', 21.9'', 14.5'', 10.6''
and 8.95'' \footnote{See web site:
http://www.submm.caltech.edu/cso/}. The pointing was checked using
planets (Jupiter, Mars and Saturn) and evolved stars (e.g. IRC 10216,
R-Hya, CRL 2688, CRL 618, NGC 7027, R-CAS and O-Ceti) for all lines
except CO(7-6). Planets were the sole pointing sources at 806 GHz.
The pointing accuracy is around 5''. The overall calibration accuracy
is $\approx$ 20\%. Data have been reduced using the GILDAS/CLASS data
analysis package. The spectra have been smoothed to a velocity
resolution of $\approx$ 10 kms$^{-1}$. Gaussian profiles have been
fitted to observed spectra (see Figs. ~\ref{fig:spec_ic10} to
~\ref{fig:spec_n6946_bras}), and linear baselines have been
subtracted.

\section{Data analysis}\label{sec:data}

    \subsection{Spectra and maps}\label{secsub:speketmap}

\begin{figure}
    \begin{center}
        \epsfxsize=9cm
        \epsfbox{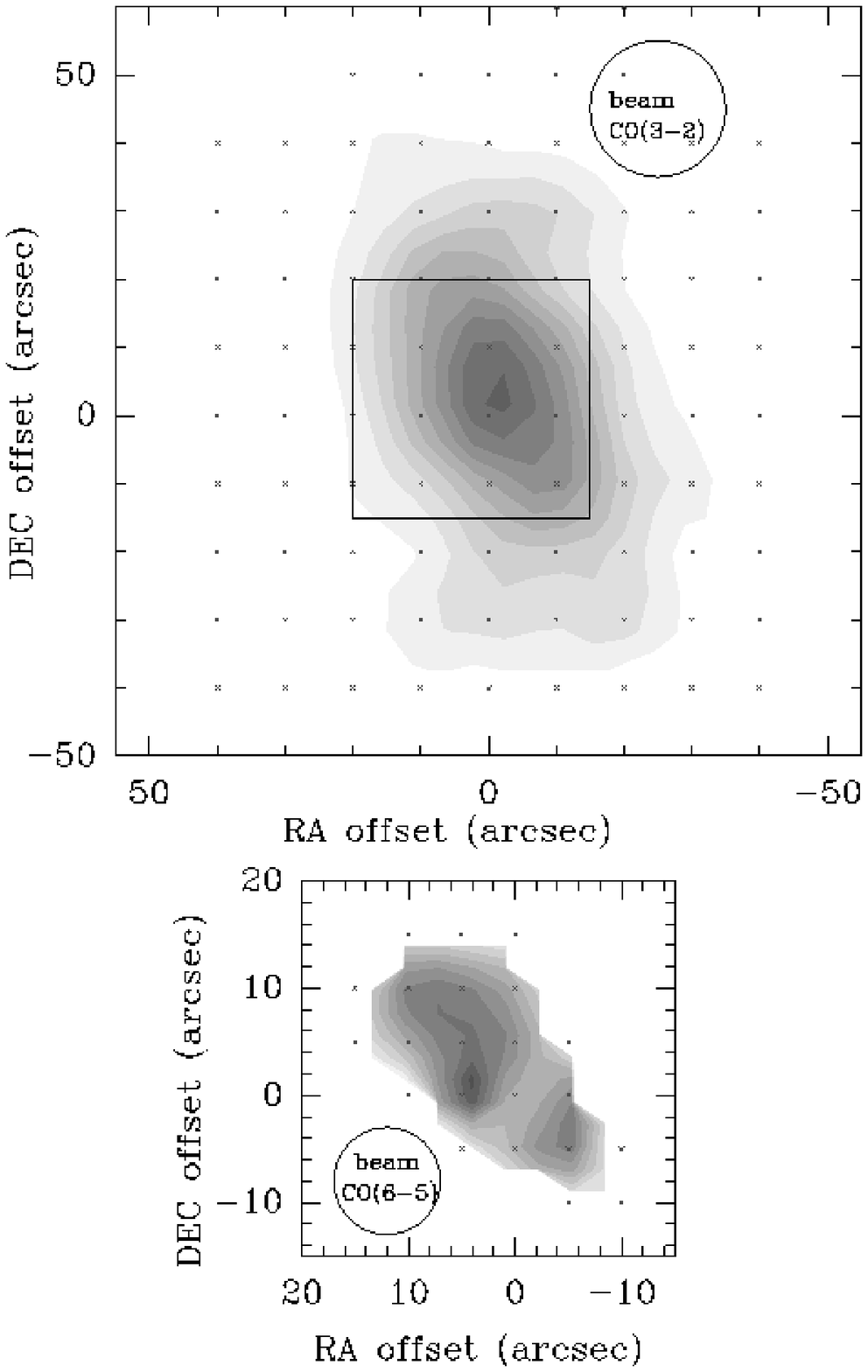}
        \caption{Map of the velocity integrated intensity of $^{12}$CO(3-2)
        (top) and $^{12}$CO(6-5) (bottom) towards IC 342. The CO emission is
        integrated over the velocity ranges -~200 to 200 kms$^{-1}$ and -50
        to 150 kms$^{-1}$, respectively. For both maps, the crosses
        show the observed positions. The contours of the
        $^{12}$CO(3-2) map range from $\int{T_{mb}dv}$ = 12
        Kkms$^{-1}$ to 112 Kkms$^{-1}$ with 10 Kkms$^{-1}$ steps. For
        the $^{12}$CO(6-5) map, the contours range from
        $\int{T_{mb}dv}$ = 8 Kkms$^{-1}$ to 108 Kkms$^{-1}$ with 10
        Kkms$^{-1}$ steps. The intensity peak value of the
        $^{12}$CO(3-2) map is $\int{T_{mb}dv} \approx$ 110
        Kkms$^{-1}$. The intensity peak value of the $^{12}$CO(6-5)
        map is $\int{T_{mb}dv} \approx$ 108 Kkms$^{-1}$. The black box
        in the upper figure represents the size of the lower
        figure. The circles indicate the beam sizes at the frequency
        of the $^{12}$CO(3-2) line (top) and at the frequency of the
        $^{12}$CO(6-5) line (bottom). No pointing error was detected
        for this map.}\label{fig:map_ic342}
    \end{center}
    \end{figure}

\begin{figure}
    \begin{center}
        \epsfxsize=9cm
        \epsfbox{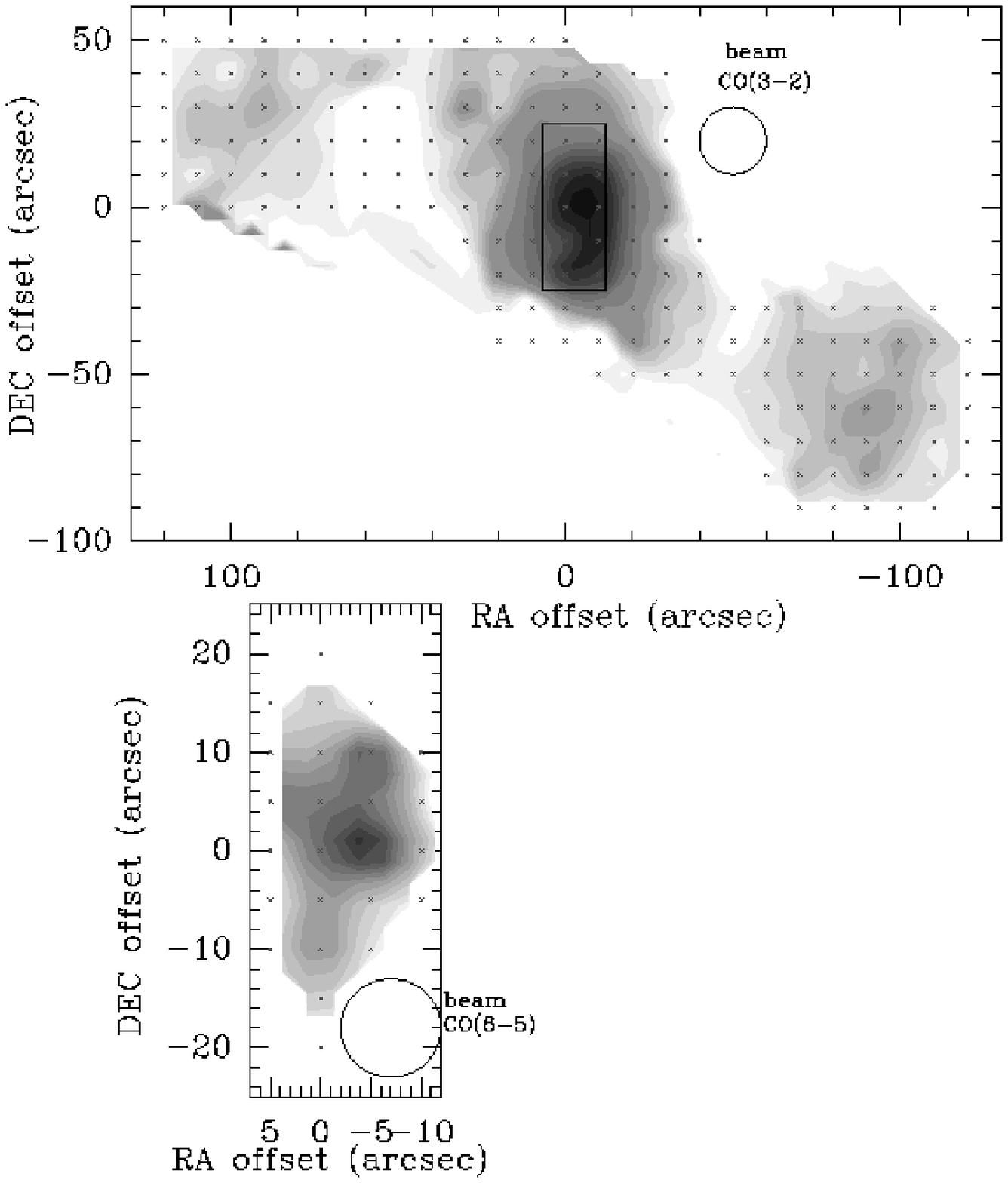}
        \caption{Map of the velocity integrated intensity of $^{12}$CO(3-2)
        (top) and $^{12}$CO(6-5) (bottom) lines towards M 83. For both
        maps, the CO emission is integrated over the velocity range
        200 to 800 kms$^{-1}$, and the crosses show the observed
        positions. The contours of the $^{12}$CO(3-2) map range from
        $\int{T_{mb}dv}$ = 3 Kkms$^{-1}$ to 18 Kkms$^{-1}$ with 3
        Kkms$^{-1}$ steps, and from $\int{T_{mb}dv}$ = 21 Kkms$^{-1}$
        to 102 Kkms$^{-1}$ with 10 Kkms$^{-1}$ steps. For the
        $^{12}$CO(6-5) map, the contours range from $\int{T_{mb}dv}$ =
        20 Kkms$^{-1}$ to 320 Kkms$^{-1}$ with 25 Kkms$^{-1}$
        steps. The intensity peak value of the $^{12}$CO(3-2) map is
        $\int{T_{mb}dv} \approx$ 100 Kkms$^{-1}$. The intensity peak
        value of the $^{12}$CO(6-5) map is $\int{T_{mb}dv} \approx$
        318 Kkms$^{-1}$. The $^{12}$CO(6-5) map has been shifted by
        +5'' along the minor axis due to a pointing error The black
        box in the upper figure represents the size of the lower
        figure. The circles indicate the beam size at the frequency of
        the $^{12}$CO(3-2) line (top) and at the frequency of the
        $^{12}$CO(6-5) line (bottom).}\label{fig:map_m83}
    \end{center}
    \end{figure}

    Spectra of the galaxy nuclei are shown in Figs. ~\ref{fig:spec_ic10}
    to ~\ref{fig:spec_n6946}. The spectra taken at positions in the
    spiral arms
    of M 83 are shown in Fig. ~\ref{fig:spec_m83_bras} while those taken
    at positions in the NGC 6946 arms are presented in
    Fig. ~\ref{fig:spec_n6946_bras}.
    The observed line is indicated above. Figs. ~\ref{fig:map_m83}
    and ~\ref{fig:map_ic342} show integrated intensity maps
    of the $^{12}$CO(3-2) (top) and the $^{12}$CO(6-5) (bottom) lines
    towards M 83 and IC 342, respectively.

    Tables ~\ref{tab:obs1} and ~\ref{tab:obs2} in appendix
    ~\ref{tables} list the line intensities (A in Kkms$^{-1}$ and I in
    Wm$^{-2}$sr$^{-1}$) and the line fluxes (F in Wm$^{-2}$) resulting
    from Gaussian fits for the sample galaxies (for each observation,
    we give the corresponding beam size). I (in Wm$^{-2}$sr$^{-1}$) is
    derived using Eqs. (3) and (4) in \citet{Baye04}. To compute the
    flux, F (in Wm$^{-2}$), we used Eq. (5) in \citet{Baye04}. The
    estimated errors on data listed in Tables ~\ref{tab:obs1} and
    ~\ref{tab:obs2} are also indicated in column 5. For most sources,
    the atomic carbon [CI]($^{3}$P$_{1}$-$^{3}$P$_{0}$) data have been
    published by \citealt{Geri00} but they have been analyzed again
    for consistency. All the CSO observations we obtained for our
    galaxy sample are summarized in Table ~\ref{tab:obs1} where
    we list the data for all positions in the nucleus as well
    as in the spiral arms when it is appropriate.

    Table ~\ref{tab:obs2} in the electronic appendix presents central
    position data for a restricted sample (IC 10, IC 342, M 83, NGC
    4038, Overlap and NGC 6946) we have been able to observe
    extensively (up to the $^{12}$CO(6-5) or $^{12}$CO(7-6) line). For
    these sources, we list the available informations in the
    literature, in the same way as \citet{Baye04} did for NGC 253 and
    Henize 2-10. All CO transitions we used for the modelling work,
    are identified with an asterisk in column 8 of Table ~\ref{tab:obs2}.

    Moreover, in order to compare at the same spatial resolution different CO
    line intensities, we have convolved the line intensities to
    a common (final) beam size of 21.9'', which is the CSO beam size at the
    frequency of the $^{12}$CO(3-2) line. Precisely, to perform this
    convolution, we multiplied A, I and F by factors depending on the
    size of each emitting source (See Eqs. ~\ref{source-ronde} to
    ~\ref{fact-ell-ge} in appendix ~\ref{cal}), the initial
    (observed) beam size and the final beam size (21.9"). To determine
    the source sizes, we used high spatial resolution maps and fitted
    Gaussian spatial profile to these maps. The sources may be either
    axisymmetric (a $\times$ a) or elliptical (a $\times$ b). Table
    ~\ref{tab:convol} lists the adopted source sizes and the maps we
    used for this measurement.

    \begin{table*}
    \caption{Source sizes used to convolve the data to 21.9". For each source,
    we used maps (specified in column 5) to determine the shape of the Gaussian
    intensity profile (axisymmetric : a $\times$ a or elliptical : a
    $\times$ b) and estimate the values of the full widths at half
    maximum (FWHM)}\label{tab:convol}
    \begin{center}
        \begin{tabular}{l c c c c}\hline\hline
        & FWHM & line (frequency) & Observed beam size & References \\
        \hline
        IC 10 & 20.0" $\times$ 20.0" & $^{12}$CO(4-3) (461 GHz) &
        14.5" & This work$^{1}$\\
        \hline
        NGC 253$^{a}$ & 23.0" $\times$ 11.0" & $^{12}$CO(6-5) (691
        GHz) & 10.6" & \citet{Baye04}$^{2}$\\
        \hline
        IC 342 & 11.0" $\times$ 11.0" & $^{12}$CO(6-5) (691 GHz) &
        10.6" & This work$^{3}$\\
        \hline
        Henize 2-10$^{a}$ & 13.0" $\times$ 13.0" & $^{12}$CO(3-2) (345
        GHz) & 22" & \citet{Meie01b}$^{4}$\\
        \hline
        NGC 4038 & 13.2" $\times$ 9.9" & $^{12}$CO(1-0) (115 GHz) &
        4.91" x 3.15" & \citet{Wils00}$^{5}$\\
        \hline
        Overlap & 11.0" $\times$ 8.8" & $^{12}$CO(1-0) (115 GHz) &
        4.91" x 3.15" & \citet{Wils00}$^{6}$\\
        \hline
        M 83 & 12.0" $\times$ 10.6" & $^{12}$CO(6-5) (691 GHz) & 10.6"
        & This work$^{7}$\\
        \hline
        NGC 6946 & 17.0" $\times$ 17.0" & $^{12}$CO(4-3) (461 GHz) &
        17" & \citet{Niet99}$^{8}$\\
        \hline
        \end{tabular}
    \end{center}
    \begin{minipage}{18cm}
    $^{a}$ : We include the sources from \citet{Baye04}. References:
    $^{1}$:The 20'' FWHM is in agreement with the size deduced from
    the CO maps found in \citet{Wils91} or in \citet{Bola00}; $^{2}$ :
    \citet{Baye04} found good agreement with the source size deduced
    from \citet{Peng96} map; $^{3}$ : The 11'' FWHM is very similar to
    the size deduced from the $^{12}$CO(4-3) map of
    \citet{Isra03}. \citet{Ecka90c} using the IRAM-30m telescope,
    obtained a map in the $^{12}$CO(2-1) line which shows a source
    size around 20''.; $^{4}$ : See \citet{Baye04}; $^{5}$ : OVRO map;
    $^{6}$ : Overlap corresponds to SGMC 4-5 in this OVRO map; $^{7}$
    : \citet{Lund04} considered the M 83 nucleus as a 12''
    axisymmetric source which agrees with our value; $^{8}$ : The
    $^{12}$CO(1-0) map published by \citet{Weli88}, is consistent with
    a 15'' source size.
    \end{minipage}
    \end{table*}

    \citet{Baye04} used the same method in their study of
    NGC 253 and Henize 2-10.

    \subsection{The C and CO cooling rates}\label{secsub:cool}

    We have derived the observed C and CO cooling rates which provide an essential
    information for the thermal balance of the studied galaxy
    nuclei. We estimated the observed C and CO cooling rates by
    summing the intensities (I in Wm$^{-2}$sr$^{-1}$) of all CO
    transitions listed in Table ~\ref{tab:obs2} and identified with an
    asterisk (both literature data and our dataset). We have computed
    the observed C and CO cooling rates in the galaxy nuclei for a
    common beam size of 21.9''. This corresponds to linear scales of
    106 pc, 191 pc, 1.5 kpc, 372 pc and 584 pc for the IC 10, IC 342,
    NGC 4038 (and the Overlap region), M 83 and NGC 6946 nuclei,
    respectively. The observed cooling rates for C and CO are listed
    in Table ~\ref{tab:resmod} together with the CO cooling rates
    derived from PDR and LVG models (see Sect.~\ref{secsub:lvg}).

    For all studied galaxies, the CO lines contributing the most to the
    observed CO cooling rates are $^{12}$CO(6-5) and $^{12}$CO(7-6),
    followed by $^{12}$CO(4-3) and $^{12}$CO(3-2), with varying relative
    contributions of those lines in the studied sources. These results
    confirm that the CO transitions with high-J (J$_{upper}$ $\geqslant$
    3) are contributing the most to the total observed CO cooling rate,
    with the highest contribution for the $^{12}$CO(6-5) line in the
    observed galaxies.

    For all targets, it is also noticeable that the observed CO
    cooling rate is higher than the observed C cooling rate, by a
    factor of $\lesssim 4.0$ for IC 10 and IC 342, of 6.9 and 19.7 for
    NGC 253 and Henize 2-10, respectively (see \citealt{Baye04} for
    these last two sources), of $\gtrsim$20 for NGC 4038, Overlap, M
    83 and NGC 6946, respectively. In the NGC 4038, Overlap, M 83 and
    NGC 6946 nuclei, the line which contributes the most to the C
    cooling rate ([CI]($^{3}$P$_{2}$-$^{3}$P$_{1}$), as it is shown in
    the following) has not been observed yet, but the difference
    between the C and CO cooling rates is large enough not to modify
    the dominant role of carbon monoxide for the gas cooling with
    respect to the atomic carbon.

    Similar results have been observed for the distant galaxies
    J1148+5251 (z=6.42) and PSS2322+1944 (z=4.12) (See \citealt{Cox02,
    Bert03, Pety04, Walt04}). The lines which contribute the most to the CO
    cooling rates are those with J$_{upper}$$\geqslant$ 3. The difference
    between the C and CO cooling rates seems also to be important (factor
    of $>$ 10) in distant objects.

    Because the observed C and CO cooling rates are computed with the
    observed lines only (see asterisks in Table ~\ref{tab:obs2}),
    unobserved lines (e.g. [CI]($^{3}$P$_{2}$-$^{3}$P$_{1}$),
    $^{12}$CO(5-4) or $^{12}$CO(8-7)) might contribute significantly
    to the CO or C cooling rates. Therefore, we used models in
    Sect. ~\ref{sec:mod} to predict the missing line intensities; and
    also to obtain the physical properties of the warm and dense gas
    contained in the galaxy nuclei. Table ~\ref{tab:resmod} shows that
    the observed CO cooling rate is $\approx$ 70\% of the modelled CO
    cooling rate for galaxies with $^{12}$CO(7-6) detections and
    25\%-50\% for galaxies without $^{12}$CO(7-6) detections.

\section{CO models}\label{sec:mod}

    \subsection{Description of the models}\label{secsub:mod-des}

    In this section, we used the measured CO line ratios from the line
    intensities (I and A) to determine the physical conditions of warm
    and dense molecular gas in galactic nuclei; namely the kinetic
    temperature (T$_{k}$), the gas density (n(H$_{2}$)), the CO column
    density divided by the line width (N(CO)/$\Delta v$) and the Far
    UV radiation field ($\chi_{FUV}$). In the first section, we use a
    LVG radiative transfer model and in Sect.~\ref{secsub:pdr}, we
    discuss the use of a PDR model. XDR model results applied to AGN
    nuclei from our sample will be presented in a forthcoming paper.

    The radiative transfer models, based on the LVG formalism, have been
    developed by \citet{Gold74, DeJo75}. The source is modelled as a one
    component spherical cloud, with uniform kinetic temperature and
    density. When using both $^{12}$CO and $^{13}$CO data, there are
    four main variables in LVG models : the molecular hydrogen density
    n(H$_2$), the gas kinetic temperature $T_K$, the CO column density
    divided by the line width, and the $^{12}$CO/$^{13}$CO abundance
    ratio. The LVG approximation is used for efficiently solving
    the radiative transfer equation, when the molecule level populations
    are not thermalized. LVG models are useful for a first order determination
    of the gas properties. However, as the medium is assumed
    to be homogeneous, these models provide an average
    description of the molecular gas, which is known to exhibit structure
    at smaller spatial scales than sampled by these observations.

    Photo-dissociation region (PDR) models are more sophisticated
    than LVG models as they solve simultaneously for the gas chemistry,
    photo-dissociation and thermal balance, taking into account
    the relevant physical and chemical processes. Such
    models have been developed during the past two decades, for a variety
    of astrophysical sources, from giant molecular clouds illuminated by
    the interstellar radiation field to the conditions experienced by
    circumstellar disks or by dense matter, very close to hot massive
    stars \citep{Tiel85a, Tiel85b, LeBo93, Koes94, Ster95, Holl99,
    Kauf99}. In PDR models, the line emission of the most
    abundant species, which determine the gas cooling rate, is usually
    calculated under the LVG approximation. Because of the
    many physical and chemical processes involved, PDR models use a simple
    geometry for the modelled cloud, which can be either plane parallel
    or spherical. When applied to CO line emission, PDR model predictions
    correspond to the integral along the line of sight of the contributions
    of the different regions in the modelled, plane parallel, cloud, with
    pronounced kinetic temperature and abundance variations from the
    warm outer shells to the cold interior. Therefore,
    the contributions of these different regions to the CO emission
    depends on the rotational line, the high-J CO lines being more
    sensitive to the warm gas ($\gtrsim$20 K), while the CO(1-0)
    contribution is biased to the cold gas ($<$20 K).

    We have chosen to use both types of models as they are built
    with different hypotheses and therefore provide different information when the model outputs are compared
    with CO observations. As LVG models are simple and fast, they are
    largely used in the analysis of extragalactic CO data. But LVG
    models are obviously a crude approximation to the complex mixture
    of physical conditions in galaxy nuclei. Although solving the
    radiative transfer under the same LVG approximation, PDR models
    consider kinetic temperature and CO abundance gradients along the
    line of sight. By comparing the PDR and LVG model predictions for
    the same source, we gain some insight on the quality of the
    models, especially on the reliability of the predictions for the
    missing CO lines, and on the computation of the CO cooling rate.

    \subsection{LVG models}\label{secsub:lvg}

        \subsubsection{Fitting procedures}\label{secsubsub:lvgfit}

        We have shown in \citet{Baye04} (Fig. 1) that the
        CO lines with J$_{upper}$ $\geqslant 3$ provide a good signature of
        the warm molecular gas ($T_{k}\gtrsim$ 20 K). In this paper,
        we used the same method, and constrain the fits of LVG or
        PDR models by the line intensity ratios from the
        observed CO submillimeter lines.

        More precisely, for IC 10 and IC 342, we
        used the following line intensity ratios (See Table
        ~\ref{tab:raplvg}) : $\frac{^{12}CO(3-2)}{^{12}CO(4-3)}$,
        $\frac{^{12}CO(3-2)}{^{12}CO(6-5)}$,
        $\frac{^{12}CO(3-2)}{^{12}CO(7-6)}$,
        $\frac{^{12}CO(2-1)}{^{12}CO(7-6)}$ and
        $\frac{^{12}CO(3-2)}{^{13}CO(3-2)}$. Some observations of NGC
        4038 and Overlap suffer from large error bars. Indeed, the
        $^{12}$CO(4-3) and the $^{12}$CO(7-6) lines have a lower
        signal-to-noise ratio than other transitions (See
        Figs.~\ref{fig:pred_Overlap} and ~\ref{fig:pred_n4038} where
        these lines appear clearly too faint). Therefore, we chose for
        those sources the following ratios to constrain the models:
        $\frac{^{12}CO(3-2)}{^{12}CO(6-5)}$,
        $\frac{^{12}CO(2-1)}{^{12}CO(6-5)}$ and
        $\frac{^{12}CO(3-2)}{^{13}CO(3-2)}$ (See Table
        ~\ref{tab:raplvg}). Since we missed the $^{12}$CO(7-6) line
        for M 83 and NGC 6946, we used the following line intensity
        ratios : $\frac{^{12}CO(3-2)}{^{12}CO(4-3)}$,
        $\frac{^{12}CO(3-2)}{^{12}CO(6-5)}$,
        $\frac{^{12}CO(2-1)}{^{12}CO(6-5)}$ and
        $\frac{^{12}CO(3-2)}{^{13}CO(3-2)}$ (See Table
        ~\ref{tab:raplvg}).

        To compute these line intensity ratios, we
        used data identified with an asterisk in Table ~\ref{tab:obs2}
        (precisely, in the column 5 : the line area, A, in
        Kkms$^{-1}$) which have been previously scaled to a common beam size
        of 21.9''(see Sect.~\ref{secsub:speketmap}). Values of line
        intensity ratios are listed in Table ~\ref{tab:raplvg}.
        The $^{13}$CO data are particularly useful for
        measuring the CO column densities while the $^{12}$CO data provide
        constraints on the kinetic temperature and the H$_2$
        density. The $^{12}$CO and $^{13}$CO spectra used are
        displayed in Figs.~\ref{fig:spec_ic10} to
        ~\ref{fig:spec_n6946} (See Sect. ~\ref{secsub:speketmap}).

        \begin{table*}
            \caption{Parameters of the ``best fit'' LVG and PDR models
            for IC 10, IC 342, NGC 4038, Overlap, M 83 and NGC
            6946. For each galaxy, the same abundance ratio
            $\frac{^{12}C}{^{13}C}^{b}$ and the same $\Delta v^{a}$
            are used for constraining both LVG and PDR models. To
            compare observations and predictions deduced from models,
            the observed CO ($^{12}$CO and $^{13}$CO) and the observed
            C cooling rates are listed at the
            bottom.}\label{tab:resmod}
            \begin{center}
                \begin{tabular}{|c|c|c|c|c|c|c|}
                \hline
                & \textbf{IC 10} & \textbf{IC 342} &  \textbf{NGC
                4038} & \textbf{Overlap} & \textbf{M 83} & \textbf{NGC 6946}\\
                \hline
                $\frac{^{12}C}{^{13}C}^{b}$ & 40 & 40 & 40 & 40 & 40 & 40\\
                \hline
                $\Delta v^{a}$ (kms$^{-1}$) & 15 & 54 & 92 & 155 & 91 & 158 \\
                \hline
                \hline
                LVG MODELS &&&&&&\\
                \hline
                N($^{12}$CO)/$\Delta v^{a}$ & 3.0$\times 10^{16}$ &
                3.5$\times 10^{17}$ & 3.2$\times 10^{16}$ & 2.4$\times
                10^{16}$ & 6.0$\times 10^{16}$ & 7.0$\times 10^{16}$\\
                (cm$^{-2}$/kms$^{-1}$) &&&&&&\\
                \hline
                n(H$_{2}$) (cm$^{-3}$) & $7.0\times 10^{5}$ &
                $1.9\times 10^{3}$ & 3.5$\times 10^{5}$ & 8.0$\times
                10^{3}$ & $6.5\times 10^{5}$ & 1.5$\times 10^{3}$ \\
                \hline
                $T_{K}$ (K) & 25 & 40 & 40 & 145 & 40 & 130 \\
                \hline
                FF$_{LVG}$$^{c}$ & $4.6\times 10^{-2}$ & $6.7\times
                10^{-2}$ & $1.7\times 10^{-2}$ & $1.3\times 10^{-2}$ &
                $3.6\times 10^{-2}$ & $2.7\times 10^{-2}$ \\
                \hline
                LVG Predicted &&&&&&\\
                $^{12}$CO cooling & $5.7\times 10^{-9}$ & $3.7\times
                10^{-8}$ & $4.1\times 10^{-8}$ & $4.3\times 10^{-8}$ &
                $1.3\times 10^{-7}$ & $4.8\times 10^{-8}$ \\
                (Wm$^{-2}$sr$^{-1}$) &&&&&&\\
                \hline
                LVG Predicted &&&&&&\\
                $^{13}$CO cooling & $2.2\times 10^{-10}$ & $1.9\times
                10^{-9}$ &  $1.0\times 10^{-9}$ & $7.4\times 10^{-10}$
                & $4.1\times 10^{-9}$ & $1.0\times 10^{-9}$ \\
                (Wm$^{-2}$sr$^{-1}$) &&&&&&\\
                \hline
                \hline
                PDR MODELS &&&&&&\\
                \hline
                FUV radiation $\chi_{FUV}$ & $5.5\times 10^{4}$ &
                $8.5\times 10^{3}$ & $2.5\times 10^{5}$ & $1.5\times
                10^{5}$ & $5.5\times 10^{4}$ & $1.5\times 10^{5}$ \\
                ($\times G_{0}$) & & & & & &\\
                \hline
                n(H) (cm$^{-3}$) & $2.0\times 10^{5}$ & $1.5\times 10^{5}$ &
                $3.0\times 10^{5}$ & $3.5\times 10^{5}$ & $5.0\times 10^{5}$ &
                $1.0\times 10^{5}$ \\
                \hline
                FF$_{PDR}$$^{c}$ & $1.3\times 10^{-2}$ & $2.9\times 10^{-2}$ &
                $9.4\times 10^{-1}$ & $9.2\times 10^{-1}$ & $1.2\times
                10^{-2}$ & $2.9\times 10^{-2}$ \\
                \hline
                PDR Predicted &&&&&&\\
                $^{12}$CO cooling & $5.8\times 10^{-9}$ & $3.6\times
                10^{-8}$ & $3.9\times 10^{-8}$ & $4.9\times 10^{-8}$ &
                $1.2\times 10^{-7}$ & $4.6\times 10^{-8}$ \\
                (Wm$^{-2}$sr$^{-1}$) &&&&&&\\
                \hline
                PDR Predicted &&&&&&\\
                $^{13}$CO cooling & $3.1\times 10^{-10}$ & $3.2\times
                10^{-9}$ & $1.2\times 10^{-9}$ & $1.8\times 10^{-9}$ &
                $5.1\times 10^{-9}$ & $2.1\times 10^{-9}$ \\
                (Wm$^{-2}$sr$^{-1}$) &&&&&&\\
                \hline
                PDR Predicted &&&&&&\\
                C cooling & $4.5\times 10^{-9}$ & $2.8\times 10^{-8}$
                & $2.9\times 10^{-8}$ & $2.6\times 10^{-8}$ &
                $3.7\times 10^{-8}$ & $8.7\times 10^{-8}$ \\
               (Wm$^{-2}$sr$^{-1}$) &&&&&&\\
                \hline
                \hline
                Observations$^{e}$ &&&&&&\\
                \hline
                Observed &&&&&&\\
                $^{12}$CO cooling & $4.0\times 10^{-9}$ & $2.5\times
                10^{-8}$ & $1.7\times 10^{-8}$ & $1.9\times 10^{-8}$ &
                $4.3\times 10^{-8}$ & $2.7\times 10^{-8}$ \\
                (Wm$^{-2}$sr$^{-1}$) &&&&&&\\
                \hline
                Observed &&&&&&\\
                $^{13}$CO cooling & $5.7\times 10^{-11}$ & $9.0\times
                10^{-10}$ & $9.9\times 10^{-11}$ & $1.2\times
                10^{-10}$ & $4.2\times 10^{-10}$ & $3.9\times
                10^{-10}$ \\
                (Wm$^{-2}$sr$^{-1}$) &&&&&&\\
                \hline
                Observed &&&&&&\\
                C cooling & $1.1\times 10^{-9}$ & $6.4\times 10^{-9}$
                & $9.2\times 10^{-10}$$^{d}$ & $7.1\times
                10^{-10}$$^{d}$ & $2.3\times 10^{-9}$$^{d}$ &
                $1.4\times 10^{-9}$$^{d}$ \\
                (Wm$^{-2}$sr$^{-1}$) &&&&&&\\
                \hline
                \end{tabular}
            \end{center}
            $^{a}$ : deduced from Gaussian fits to the $^{12}$CO(3-2) line
            profiles. $^{b}$ : Assumed $\frac{^{12}C}{^{13}C}$. $^{c}$ : FF
            is the filling factor of the CO emitting region in the 21.9''
            beam. $^{d}$ : Based on the sole C($^{3}$P$_{1}$-$^{3}$P$_{0}$)
            transition. $^{e}$ : See C and CO lines identified with an
            asterisk in Table ~\ref{tab:obs2}.
        \end{table*}

        \begin{table*}
            \caption{Observed and predicted (``best'' LVG model) line
            ratios for the studied sources.}\label{tab:raplvg}
            \begin{center}
                \begin{tabular}{|c|c|c|c|c|c|c|c|c|}
                \hline
                A & IC 10 & IC 342 & A & NGC 4038 & Overlap & A & M 83
                & NGC 6946\\
                (Kkms$^{-1}$)& obs.$^{*}$ & obs.$^{*}$ & (Kkms$^{-1}$)
                & obs.$^{*}$ & obs.$^{*}$ & (Kkms$^{-1}$) & obs.$^{*}$
                & obs.$^{*}$ \\
                \hline
                $\frac{^{12}CO(3-2)}{^{12}CO(4-3)}$ &
                1.8$\pm$0.3$^{a}$ & 2.2$\pm$0.3$^{a}$ &
                $\frac{^{12}CO(3-2)}{^{12}CO(4-3)}$ & 1.5$\pm$0.2 &
                2.4$\pm$0.3 & $\frac{^{12}CO(3-2)}{^{12}CO(4-3)}$ &
                1.1$\pm$0.4$^{a}$ & 1.5$\pm$0.3$^{a}$\\
                \hline
                $\frac{^{12}CO(3-2)}{^{12}CO(6-5)}$ &
                2.5$\pm$0.5$^{a}$ & 4.1$\pm$0.4$^{a}$ &
                $\frac{^{12}CO(3-2)}{^{12}CO(6-5)}$ &
                1.1$\pm$0.2$^{a}$ & 1.7$\pm$0.8$^{a}$ &
                $\frac{^{12}CO(3-2)}{^{12}CO(6-5)}$ &
                1.2$\pm$0.3$^{a}$ & 4.6$\pm$1.0$^{a}$\\
                \hline
                $\frac{^{12}CO(3-2)}{^{12}CO(7-6)}$ &
                8.7$\pm$2.4$^{a}$ & 15.0$\pm$1.6$^{a}$ & - & - & - & -
                & - & - \\
                \hline
                $\frac{^{12}CO(2-1)}{^{12}CO(7-6)}$ &
                10.8$\pm$4.6$^{a}$ & 18.5$\pm$5.3$^{a}$ &
                $\frac{^{12}CO(2-1)}{^{12}CO(6-5)}$ &
                1.6$\pm$0.5$^{a}$ & 2.2$\pm$1.3$^{a}$ &
                $\frac{^{12}CO(2-1)}{^{12}CO(6-5)}$ &
                1.4$\pm$0.1$^{a}$ & 6.9$\pm$1.6$^{a}$ \\
                \hline
                \hline
                $\frac{^{12}CO(3-2)}{^{13}CO(3-2)}$ &
                10.3$\pm$2.1$^{a}$ & 5.5$\pm$0.7$^{a}$ &
                $\frac{^{12}CO(3-2)}{^{13}CO(3-2)}$ &
                18.0$\pm$11.0$^{a}$ & 30.2$\pm$8.2$^{a,b}$ &
                $\frac{^{12}CO(3-2)}{^{13}CO(3-2)}$ &
                8.5$\pm$2.6$^{a}$ & 17.1$\pm$2.0$^{a}$\\
                \hline
                $\frac{^{12}CO(1-0)}{^{13}CO(1-0)}$ & - & 11.3$\pm$0.4 &
                $\frac{^{12}CO(1-0)}{^{13}CO(1-0)}$ & - & - &
                $\frac{^{12}CO(1-0)}{^{13}CO(1-0)}$ & - & 11.1$\pm$4.4\\
                \hline
                $\frac{^{12}CO(2-1)}{^{13}CO(2-1)}$ & 15.4$\pm$6.2 &
                10.9$\pm$2.6 & $\frac{^{12}CO(2-1)}{^{13}CO(2-1)}$ &
                24.3$\pm$9.7 & 16.3$\pm$6.6 &
                $\frac{^{12}CO(2-1)}{^{13}CO(2-1)}$ & 4.5$\pm$0.5 &
                29.7$\pm$6.7 \\
                \hline
                 & LVG model & LVG model & & LVG model & LVG model &
                & LVG model & LVG model\\
                 & T$_{K}=$25 K & T$_{K}=$40 K &  & T$_{K}=$40 K &
                T$_{K}=$145 K &  & T$_{K}=$40 K & T$_{K}=$130 K\\
                \hline
                $\frac{^{12}CO(3-2)}{^{12}CO(4-3)}$ & 1.2 & 1.3 &
                $\frac{^{12}CO(3-2)}{^{12}CO(4-3)}$ & 1.1 & 1.1 &
                $\frac{^{12}CO(3-2)}{^{12}CO(4-3)}$ & 1.1 & 1.4\\
                \hline
                $\frac{^{12}CO(3-2)}{^{12}CO(6-5)}$ & 2.8 & 4.1 &
                $\frac{^{12}CO(3-2)}{^{12}CO(6-5)}$ & 1.6 & 2.7 &
                $\frac{^{12}CO(3-2)}{^{12}CO(6-5)}$ & 1.4 & 5.1\\
                \hline
                $\frac{^{12}CO(3-2)}{^{12}CO(7-6)}$ & 8.4 & 15.6 &
                $\frac{^{12}CO(3-2)}{^{12}CO(7-6)}$ & 2.7 & 5.9 &
                $\frac{^{12}CO(3-2)}{^{12}CO(7-6)}$ & 1.9 & 13.9\\
                \hline
                $\frac{^{12}CO(2-1)}{^{12}CO(7-6)}$ & 9.1 & 17.9 &
                $\frac{^{12}CO(2-1)}{^{12}CO(6-5)}$ & 1.5 & 2.3 &
                $\frac{^{12}CO(2-1)}{^{12}CO(6-5)}$ & 1.4 & 6.1\\
                \hline
                \hline
                $\frac{^{12}CO(3-2)}{^{13}CO(3-2)}$ & 11.6 & 5.4 &
                $\frac{^{12}CO(3-2)}{^{13}CO(3-2)}$ & 16.7 & 23.9 &
                $\frac{^{12}CO(3-2)}{^{13}CO(3-2)}$ & 10.1 & 17.1\\
                \hline
                $\frac{^{12}CO(1-0)}{^{13}CO(1-0)}$ & 27.3 & 4.2 &
                $\frac{^{12}CO(1-0)}{^{13}CO(1-0)}$ & 33.7 & 36.8 &
                $\frac{^{12}CO(1-0)}{^{13}CO(1-0)}$ & 28.6 & 18.6\\
                \hline
                $\frac{^{12}CO(2-1)}{^{13}CO(2-1)}$ & 14.4 & 3.5 &
                $\frac{^{12}CO(2-1)}{^{13}CO(2-1)}$ & 22.3 & 27.9 &
                $\frac{^{12}CO(2-1)}{^{13}CO(2-1)}$ & 14.8 & 12.7\\
                \hline
                \end{tabular}
            \end{center}
            $^{*}$ : ratio derived from observations marked with
            asterisks in Table ~\ref{tab:obs2}; $^{a}$ : values used
            as constraints for the LVG models; $^{b}$ : value from
            Table 2 in \citet{Zhu03}.\\
        \end{table*}

        We made use of a least square fitting method (taking into
        account the errors of the observed intensity line ratios) to
        determine the physical conditions which reproduce the
        observations the best. The line width $\Delta v$ and the
        $\frac{^{12}\text{CO}}{^{13}\text{CO}}$ abundance ratio
        ($X_{galaxy}$ in \citet{Baye04}), are not part of the fitting
        process. $\Delta v$ is set to the value deduced from Gaussian
        fits of the $^{12}$CO(3-2) spectra (FWHM). The $\Delta v$
        values used for the target galaxies are reported in Table
        ~\ref{tab:resmod}. The $\frac{^{12}\text{CO}}{^{13}\text{CO}}$
        abundance ratio is set to 40 (see Table ~\ref{tab:resmod}) for
        IC 10, IC 342, NGC 4038, Overlap, M 83 and NGC 6946, the value
        used in previous LVG modelling works on the same galaxies (for
        IC 342 in \citealt{Maue93, Henk98}; for NGC 4038 in
        \citealt{Zhu03}, for M 83 in \citealt{Maue93} and for NGC
        6946, in \citealt{Isra01}). But for IC 10, \citet{Peti98a}
        used a $\frac{^{12}\text{CO}}{^{13}\text{CO}}$ abundance ratio
        of 50 and for Overlap, \citet{Zhu03} used two different
        abundance ratios (40 and 60).

        \subsubsection{Results}\label{secsubsub:lvgres}

\begin{figure*}
            \begin{center}
            \hspace*{-0.5cm}
            \epsfxsize=12cm
            \epsfbox{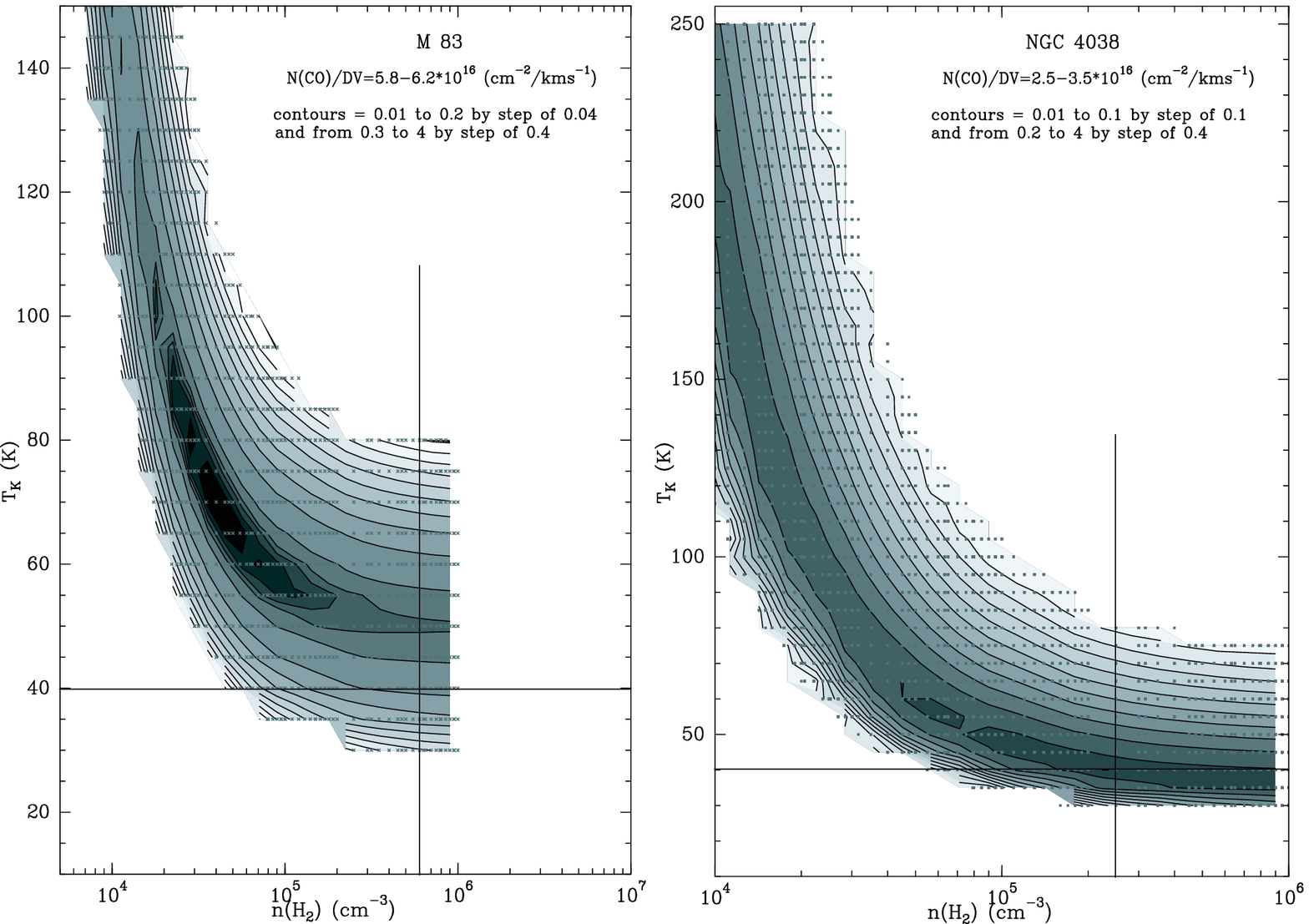}
            \vspace*{1.1cm}\caption{LVG model results for NGC 4038 (right side)
            and M 83 (left side). For each source, the ``best'' fit
            N($^{12}$CO)/$\Delta v$ (indicated at the
            top) is found by minimizing
            $\chi^{2}$. In fact, to avoid gridding problems, we use
            all models within a small interval centered around the best
            N($^{12}$CO)/$\Delta v$ value, listed in Table ~\ref{tab:resmod}.
            We plot $\Delta \chi^{2}$ contours in the $T_{K}$ (K) vs.
            n(H$_{2}$)($cm^{-3}$) space. The contour levels for M 83
            and for NGC 4038 are indicated in the plots. The LVG
            models with higher $\Delta \chi^{2}$ ($>$4)
            are not represented. In both figures, the
            dark areas represent the zones where $\chi^{2}$ is the
            lowest. Each cross corresponds to a LVG model solution.
            The best fit is located at the intersection of the two
            black lines.}\label{fig:cont_deltakhi2_ant2M83}
            \end{center}
        \end{figure*}

        For these six sources, we varied N($^{12}$CO) from 1.0$\times
        10^{16}$ cm$^{-2}$ to 1.0$\times 10^{20}$ cm$^{-2}$, T$_{K}$
        from 10 K to 255 K and n(H$_{2}$) from 10 cm$^{-3}$ to
        $10^{7}$ cm$^{-3}$. Model solutions (physical parameters which
        reproduce the observations the best) for each source are given
        in Table ~\ref{tab:resmod}. As an example of the used fitting
        process, we present in Fig. ~\ref{fig:cont_deltakhi2_ant2M83}
        results of the LVG model calculations for NGC 4038 (right
        side) and M 83 (left side). In this figure, we plot $\Delta
        \chi^{2}$ ($\chi^{2}$-$\chi^{2}_{min}$) contours in the 2D
        parameter space (n(H$_{2}$)(cm$^{-3}$) vs. $T_{K}$(K)). In
        this figure, the parameter N($^{12}$CO)/$\Delta v$ has been
        set to its ''best fit'' value. More precisely, in
        Fig. ~\ref{fig:cont_deltakhi2_ant2M83}, we have selected all
        LVG model solutions within a small interval in
        N($^{12}$CO)/$\Delta v$ centered around the best
        N($^{12}$CO)/$\Delta v$ value (see the chosen interval values
        at the top of each plot)  for a better gridding of the
        (n(H2),Tk) parameter space. Similar plots were obtained for
        each target. The ``best fit'' model (with $\chi^{2}_{min}$),
        is located at the intersection of the two black lines in
        Fig. ~\ref{fig:cont_deltakhi2_ant2M83}. Predicted line
        intensity ratios from the best fit models are listed in Table
        ~\ref{tab:raplvg}. In Fig. ~\ref{fig:cont_deltakhi2_ant2M83}
        the best models are located in the black zones which have a
        "banana" shape in the (n(H$_{2}$), $T_{K}$) plane. It is clear
        that the fitting process using the LVG framework is highly
        degenerated since acceptable fits (black zones) can be
        obtained over a large domain of the parameter space. For NGC
        4038, we conclude that the gas density is not well constrained
        while the temperature range is reasonably narrow. For M 83, we
        have the reverse case: the gas density is well constrained
        while the temperature range is broad. We notice that, despite
        the degeneracy, predicted CO line intensities computed from
        models located in the black areas in
        Figs. ~\ref{fig:cont_deltakhi2_ant2M83} do not differ
        significantly (variations in line intensity between models
        localized in the black areas are $\lesssim$ 3\%). Therefore,
        LVG predictions are relevant for computing the CO cooling
        rate. In Figs. ~\ref{fig:pred_ic10}, ~\ref{fig:pred_ic342},
        ~\ref{fig:pred_n4038}, ~\ref{fig:pred_Overlap},
        ~\ref{fig:pred_m83} and ~\ref{fig:pred_n6946}, we present the
        best LVG models for the studied sources (IC 10, IC 342, NGC
        4038, Overlap, M 83 and NGC 6946). For each source, we list
        the physical parameters corresponding to the best solutions
        in Table ~\ref{tab:resmod}. Comparisons between observed and
        modelled line intensity ratios are given in Table
        ~\ref{tab:raplvg}.

        The LVG models have been used for predicting $^{12}$CO line
        intensities from $^{12}$CO(1-0) up to $^{12}$CO(15-14) and the
        $^{13}$CO line intensities from $^{13}$CO(1-0) up to
        $^{13}$CO(6-5). The predicted $^{12}$CO and the $^{13}$CO
        cooling rates are computed by summing all $^{12}$CO and
        $^{13}$CO line intensities and are listed in Table
        ~\ref{tab:resmod}. We obtained the following results :

        \begin{itemize}
            \item [*] The ``best'' model for IC 10 is illustrated  with
            grey triangles in Fig. ~\ref{fig:pred_ic10}. The lines
            which contribute the most to the $^{12}$CO and to the
            $^{13}$CO cooling rates are the $^{12}$CO(5-4) line
            (26.6\% of the total intensity), the $^{12}$CO(6-5) line
            (25.1\%) and the $^{12}$CO(4-3) line (17.9\%), the
            $^{13}$CO(4-3) line (30.0\%), the $^{13}$CO(5-4) line
            (29.5\%) and the $^{12}$CO(6-5) line (18.3\%),
            respectively.\\

            \item [*] The ``best'' model for IC 342 is illustrated
            with grey triangles in Fig. ~\ref{fig:pred_ic342}. The
            lines which contribute the most to the $^{12}$CO cooling
            rate are $^{12}$CO(5-4) (28.1\% of the total intensity),
            $^{12}$CO(6-5) (22.2\%) and $^{12}$CO(4-3) (21.7\%) while
            the lines which dominate the $^{13}$CO cooling rate are
            $^{13}$CO(3-2) (36.3\%), $^{13}$CO(4-3) (27.7\%) and
            $^{13}$CO(2-1) (19.2\%).\\

            \item [*] For NGC 4038, we obtained a ``best'' model
            represented in Fig. ~\ref{fig:pred_n4038} with grey
            triangles. The lines which contribute the most to the
            $^{12}$CO cooling rate are $^{12}$CO(6-5) line (22.8\%),
            the $^{12}$CO(7-6) line (21.16\%) and the $^{12}$CO(5-4)
            line (17.4\%) while the main lines of the $^{13}$CO
            cooling rate are the $^{13}$CO(5-4) (33.0\%),
            $^{13}$CO(6-5) (32.3\%) and the $^{13}$CO(4-3) (22.9\%)
            lines.\\

            \item [*] For Overlap, we obtained a ``best'' model
            represented in Fig. ~\ref{fig:pred_Overlap} with grey
            triangles. The $^{12}$CO(6-5) line represents 21.2\% of
            the $^{12}$CO cooling rate while the $^{12}$CO(5-4) and
            the $^{12}$CO(7-6) lines correspond to 21.0\% and 15.7\%,
            respectively. The main lines for the $^{13}$CO cooling are
            the $^{13}$CO(5-4) (29.9\%), $^{13}$CO(4-3) (28.3\%) and
            the $^{13}$CO(6-5) (22.9\%) lines.\\

            \item [*] For M 83, the ``best'' model is represented in
            Fig. ~\ref{fig:pred_m83} with grey triangles. The
            $^{12}$CO(7-6) line represents 22.7\% of the $^{12}$CO
            cooling rate while the $^{12}$CO(6-5) and the
            $^{12}$CO(8-7) lines correspond to 19.4\% and 18.4\%,
            respectively. The $^{12}$CO cooling rate deduced from this
            ``best'' model is less accurate than for the previous
            galaxies since we constrained LVG models without the
            $^{12}$CO(7-6) line. The $^{13}$CO cooling rate is
            dominated by the intensities of the $^{13}$CO(6-5)
            (34.0\%), $^{13}$CO(5-4) (32.8\%) and the $^{13}$CO(4-3)
            (22.0\%) lines.\\

            \item [*] For NGC 6946, we obtained a ``best'' model
            represented in Fig. ~\ref{fig:pred_n6946} with grey
            triangles. The main lines for the $^{12}$CO cooling rate
            are the $^{12}$CO(5-4) (23.6\%), the $^{12}$CO(4-3)
            (20.3\%) and the $^{12}$CO(6-5) (18.9\%) lines. The
            $^{12}$CO cooling rate deduced from this ``best'' model is
            less accurate than for the previous galaxies since we
            constrained LVG models without the $^{12}$CO(7-6)
            line. The $^{13}$CO cooling rate is dominated by the
            intensities of the $^{13}$CO(3-2) (28.8\%), $^{13}$CO(4-3)
            (27.5\%) and the $^{13}$CO(5-4) (18.4\%) lines.\\
        \end{itemize}

        We present in Figs. ~\ref{fig:pred_ic10},
        ~\ref{fig:pred_ic342}, ~\ref{fig:pred_n4038},
        ~\ref{fig:pred_Overlap}, ~\ref{fig:pred_m83} and
        ~\ref{fig:pred_n6946}, the predicted  integrated areas (A in
        Kkms$^{-1}$) and the predicted line intensities (I in
        Wm$^{-2}$ sr$^{-1}$) for all $^{12}$CO transitions up to
        J=15-14 and for all $^{13}$CO transitions up to
        $^{13}$CO(6-5). We have computed the predicted integrated
        areas, A, using the predicted antenna temperature from the
        models and multiplying it by the line width $\Delta v$ and by
        the surface filling factor, FF$_{LVG}$ (both listed in Table
        ~\ref{tab:resmod}). For each observed CO line, we have
        estimated the surface filling factor (FF$_{LVG}$) of molecular
        clouds in the 21.9'' beam using the ratio
        A$_{observations}$(Kkms$^{-1}$)/A$_{model}$(Kkms$^{-1})$. We
        have made an average, weighted by the S/N, for the filling
        factors pertaining to each observed transition. For NGC 4038
        and Overlap, we have not taken into account the $^{12}$CO(4-3)
        and the $^{12}$CO(7-6) lines because they have the lowest
        signal-to-noise ratio (see Figs. ~\ref{fig:pred_n4038} and
        ~\ref{fig:pred_Overlap}). For the studied sources, we have
        obtained surface filling factor of a few \% (1.3 to 6.7\%) in
        a 21.9" beam size.

        \subsubsection{Discussion}\label{secsubsub:lvgdis}

\begin{figure*}
            \begin{center}
            \epsfxsize=14cm
            \epsfbox{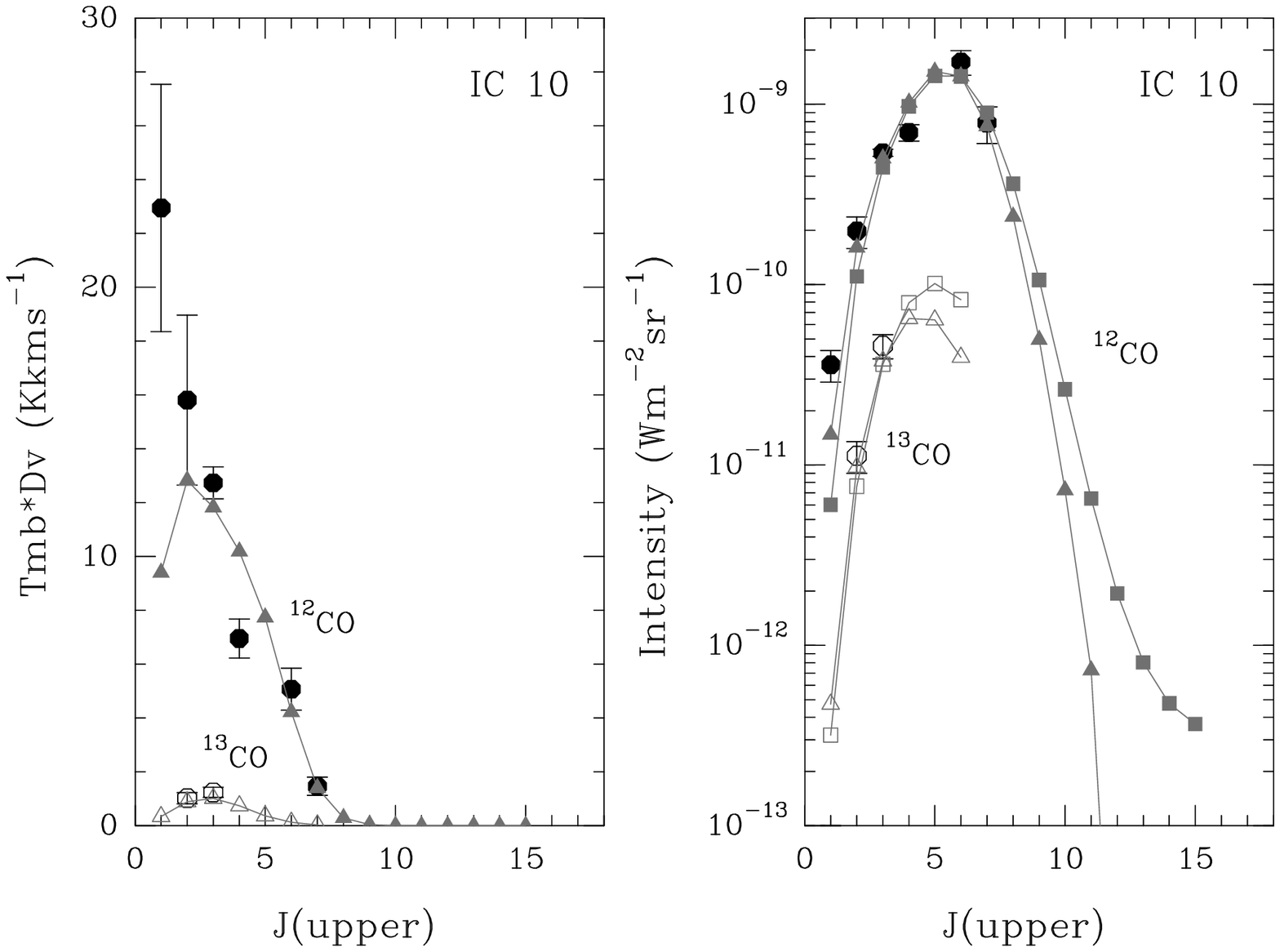}
            \caption{``best'' LVG and PDR models compared with
            observations for IC 10 (See Table ~\ref{tab:resmod} for the
            model parameters) . On the left side, we plot integrated
            intensities in Kkms$^{-1}$ vs J$_{upper}$. On the right side, we
            plot I in Wm$^{-2}$sr$^{-1}$ vs J$_{upper}$. In all figures, grey
            triangles represent the LVG model while the grey squares represent
            results from the PDR model (see
            Sect.~\ref{secsub:pdr}). Observations (with error bars) taken from
            the literature and from our data set are shown in black. The
            $^{13}$CO transitions are represented with open symbols while the
            $^{12}$CO transitions use filled symbols.}\label{fig:pred_ic10}
            \epsfxsize=14cm
            \epsfbox{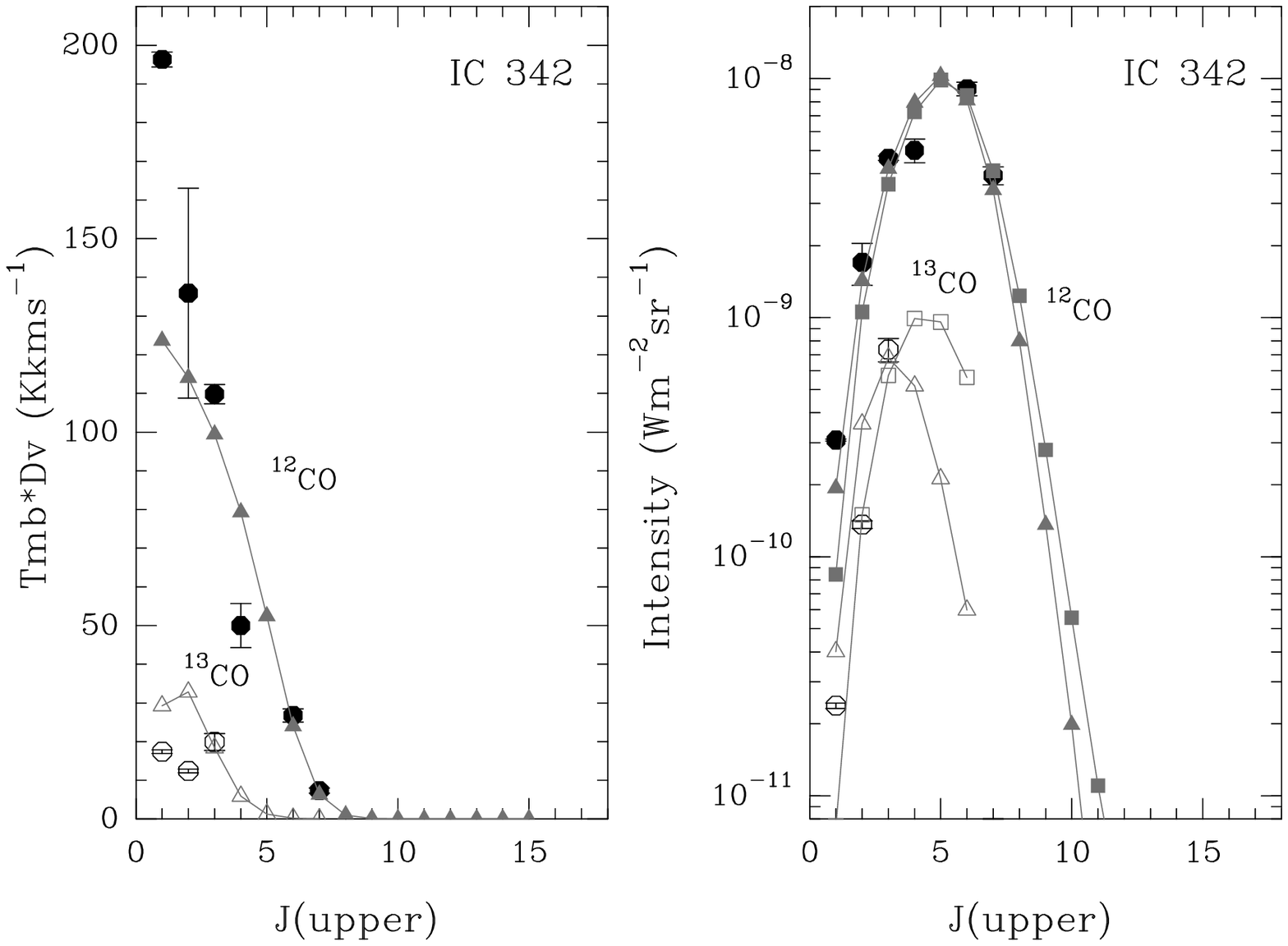}
            \caption{``best''LVG and PDR models compared with
            observations for IC 342 (See Table ~\ref{tab:resmod} for
            the model parameters). See caption of
            Fig.~\ref{fig:pred_ic10}.}\label{fig:pred_ic342}
            \end{center}
        \end{figure*}

\begin{figure*}
            \begin{center}
            \epsfxsize=14cm
            \epsfbox{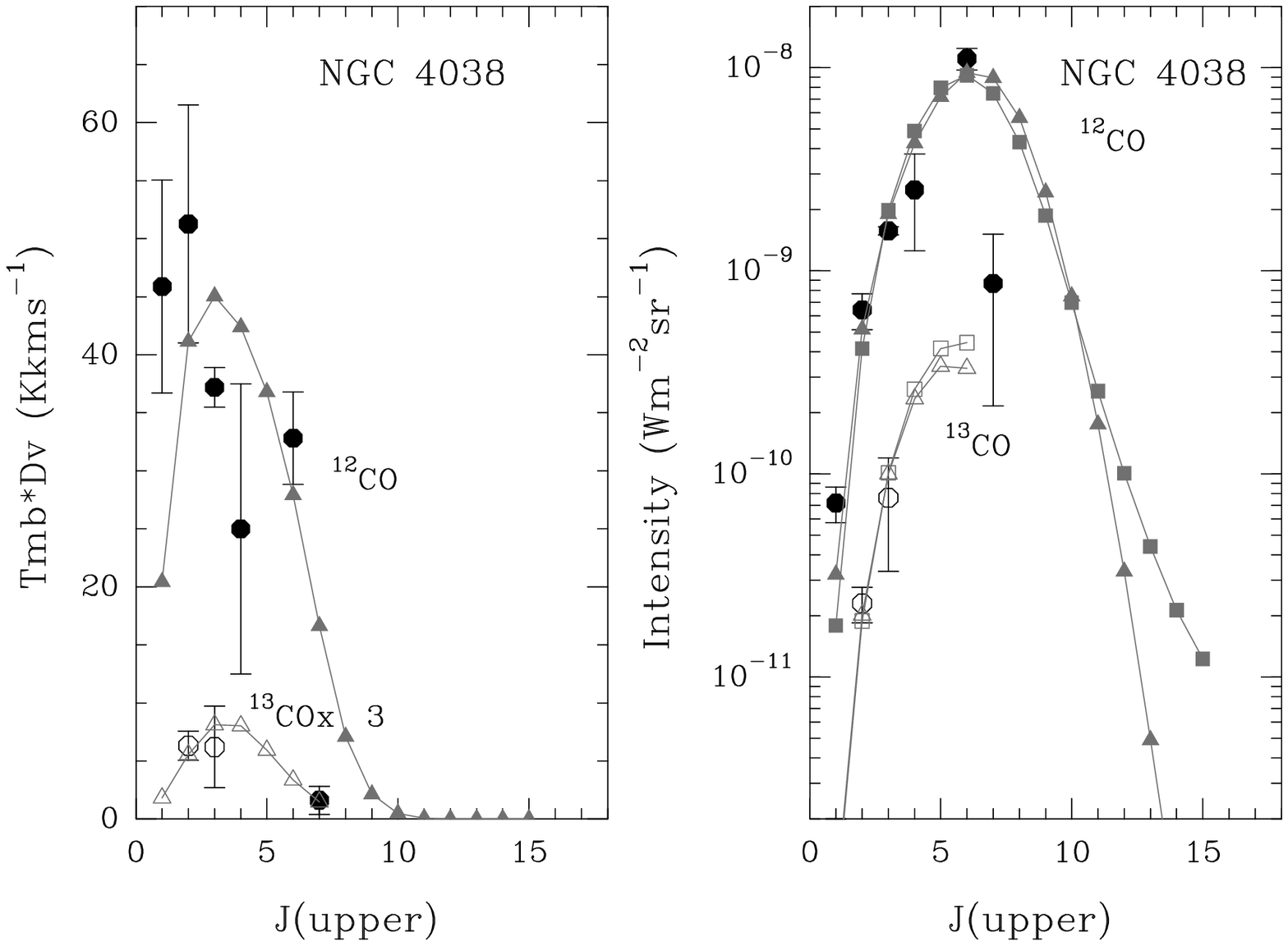}
            \caption{``best'' LVG and PDR models compared with
            observations for NGC 4038 (See Table ~\ref{tab:resmod} for
            the model parameters). See caption of
            Fig.~\ref{fig:pred_ic10}.}\label{fig:pred_n4038}
            \epsfxsize=14cm
            \epsfbox{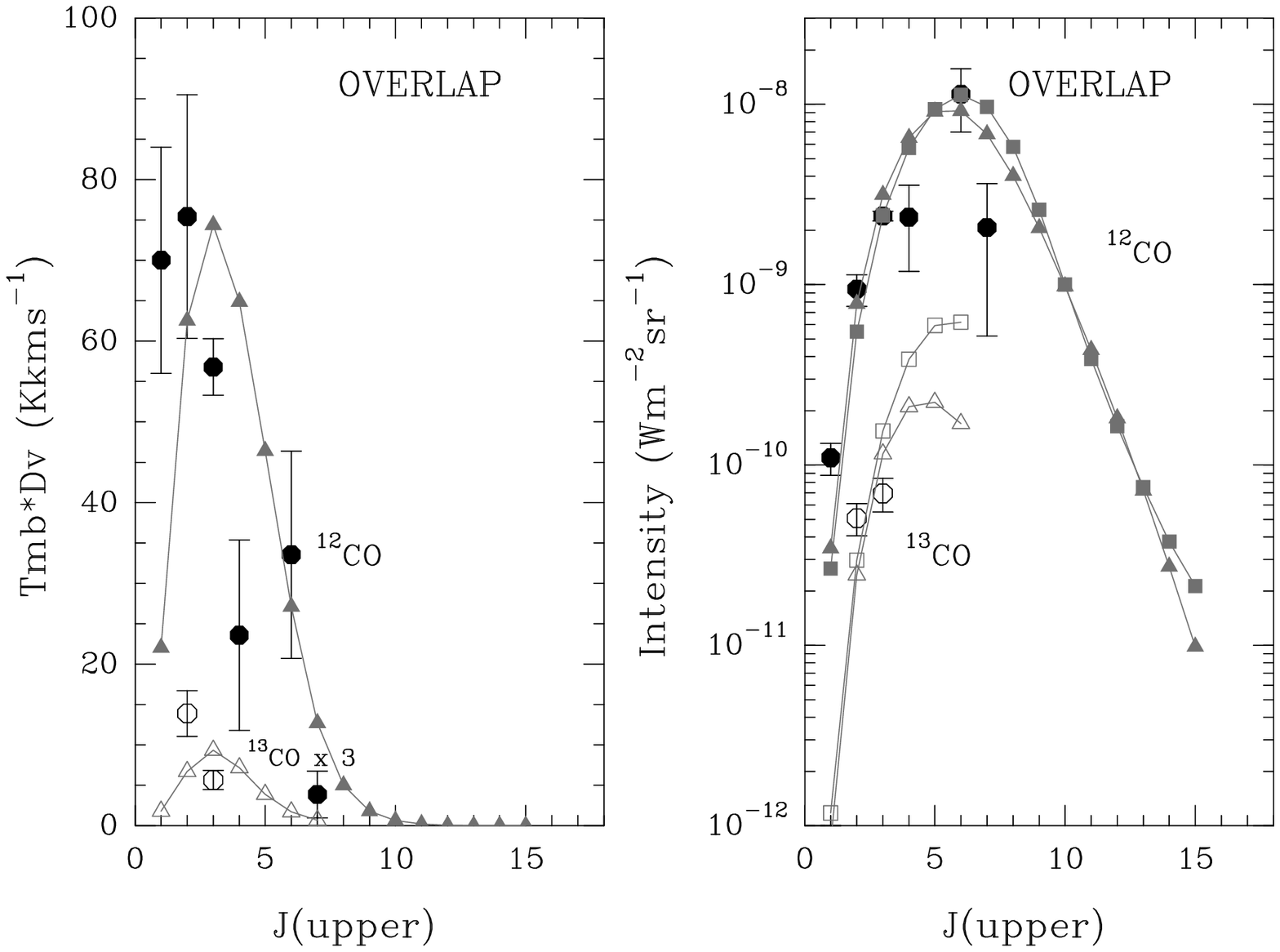}
            \caption{``best'' LVG and PDR models compared with
            observations for Overlap (See Table ~\ref{tab:resmod} for
            the model parameters). See caption of
            Fig.~\ref{fig:pred_ic10}.}\label{fig:pred_Overlap}
            \end{center}
        \end{figure*}

\begin{figure*}
            \begin{center}
            \epsfxsize=14cm
            \epsfbox{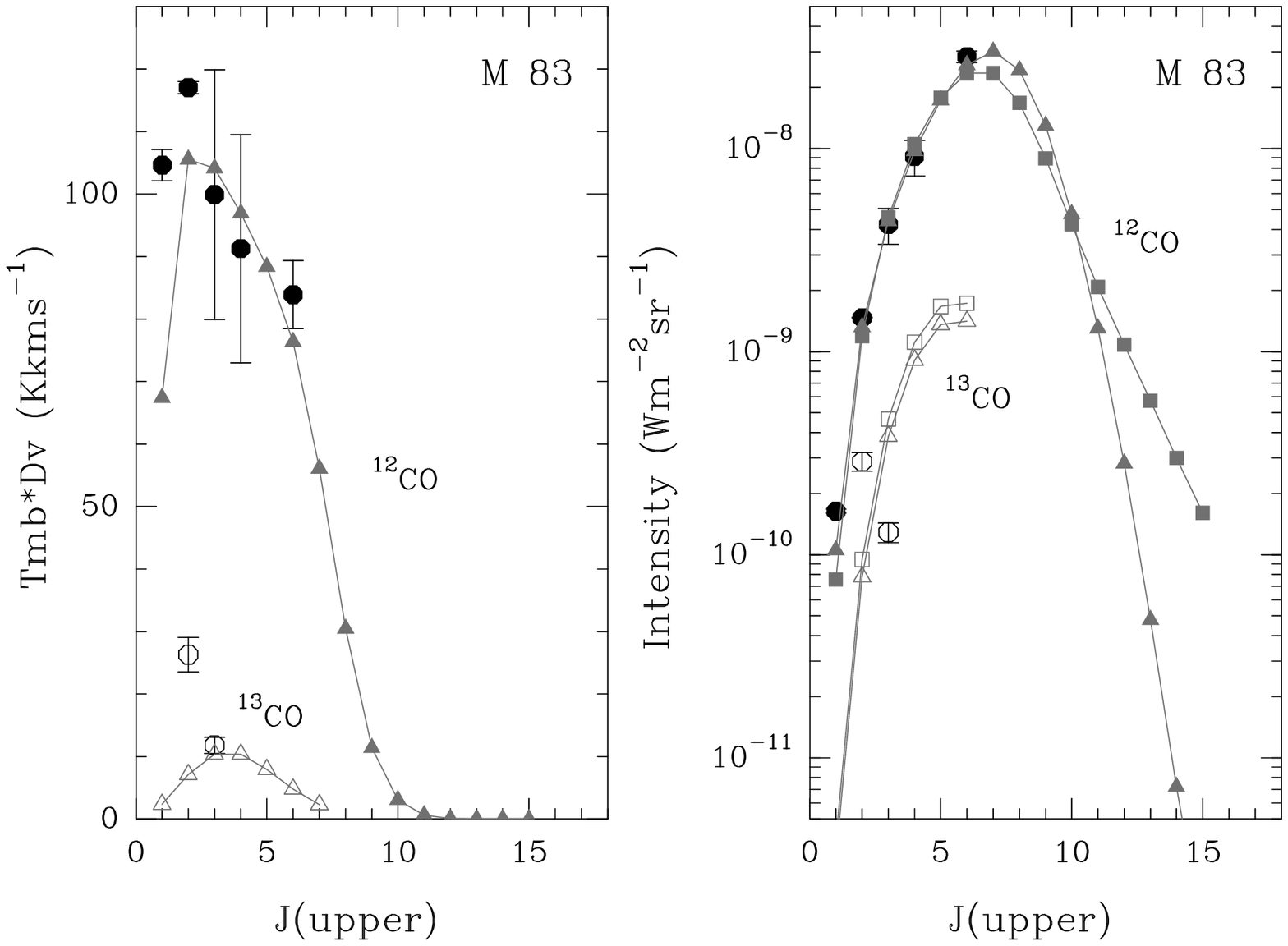}
            \caption{``best'' LVG and PDR models compared with
        observations for M 83 (See Table ~\ref{tab:resmod} for
        the model parameters).  See caption of
        Fig.~\ref{fig:pred_ic10}.}\label{fig:pred_m83}
        \epsfxsize=14cm
            \epsfbox{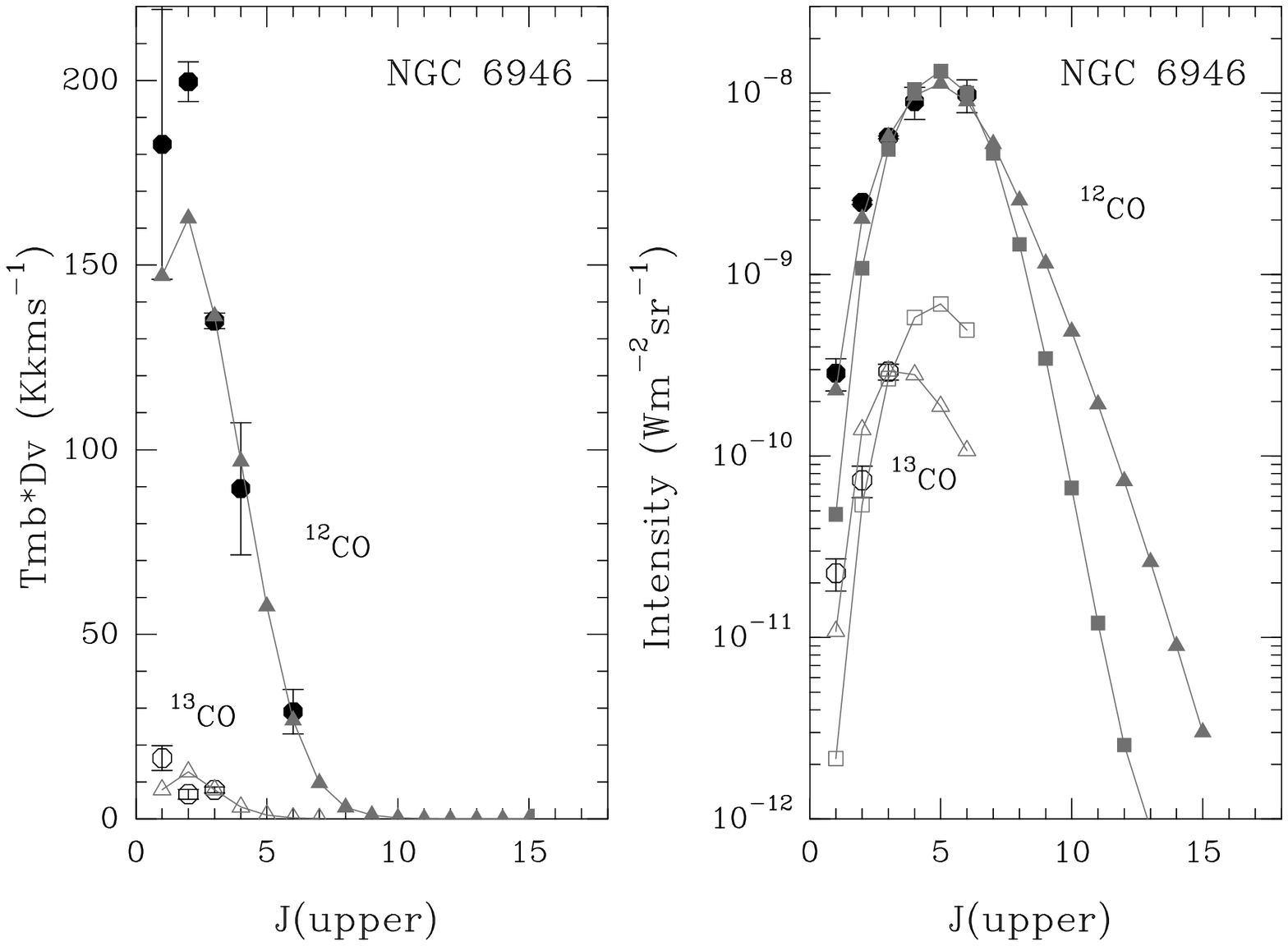}
            \caption{``best'' LVG and PDR models compared with
            observations for NGC 6946 (See Table ~\ref{tab:resmod} for
            the model parameters). See caption of
            Fig.~\ref{fig:pred_ic10}.}\label{fig:pred_n6946}
            \end{center}
        \end{figure*}

        From the above analysis we can conclude that LVG models can be
        confidently used to determine the CO cooling rates, provided that a
        sufficiently complete data set is available. Even though
        several points in the parameter space with different physical
        conditions fit the observed data equally well, the derived CO
        cooling rates are very similar in all studied cases. The key
        point is the combination of $^{12}$CO(6-5) and $^{12}$CO(7-6)
        data as these two lines provide the largest contribution to
        the observed $^{12}$CO cooling. It will be particularly
        interesting to observe the $^{12}$CO(5-4) line which is shown
        to be also a main cooling line in these galaxies. In addition
        to the $^{12}$CO(5-4), $^{12}$CO(6-5) and $^{12}$CO(7-6)
        lines, data for $^{12}$CO(8-7) and $^{12}$CO(9-8) lines would
        be most useful to discriminate models, and for a more accurate
        determination of the $^{12}$CO cooling rates. We also showed
        that $^{12}$CO lines with $J_{upper} \geqslant 10$ are
        predicted to be weak and will not have significant antenna
        temperatures (see Figs.~\ref{fig:pred_ic10},
        ~\ref{fig:pred_ic342}, ~\ref{fig:pred_n4038},
        ~\ref{fig:pred_Overlap}, ~\ref{fig:pred_m83} and
        ~\ref{fig:pred_n6946}, plots on the left side). In addition
        $^{13}$CO(6-5) data would also be extremely useful for
        constraining the models, for better estimating the $^{13}$CO
        cooling rates and for measuring the opacity of $^{12}$CO(6-5)
        line. Indeed, when comparing the predicted and observed
        intensities of the $^{13}$CO lines, the importance of high-J
        $^{13}$CO lines shows up. For the studied galaxies,
        $^{13}$CO(3-2) is the most intense observed line (see this
        work and \citealt{Baye04}). For most sources, predicted
        $^{13}$CO(6-5) line intensities are at least as strong as
        $^{13}$CO(3-2) intensities.

        Because the LVG models have been constrained by the high-J CO
        line intensity ratios, intensities of the low-J CO transitions
        are not well fitted (see for instance the $^{12}$CO(1-0) or
        the $^{12}$CO(2-1) predicted intensities in
        Figs.~\ref{fig:pred_ic10}, ~\ref{fig:pred_ic342},
        ~\ref{fig:pred_n4038} and ~\ref{fig:pred_n6946}). To relieve
        that problem, a two component LVG model would be needed, one
        component fitting the low-J CO transitions and another one for
        the high-J CO transitions. It was not our purpose here but
        \citet{Harr99, Brad03} did it for the galaxy NGC 253.

        We have compared the physical parameters (See Table
        ~\ref{tab:resmod}) corresponding to our best fit LVG models
        with results obtained in previous studies for the same
        galaxies. For IC 10, \citet{Peti98a} presented an acceptable
        LVG solution: N($^{12}$CO)/$\Delta v=$5.0$\times 10^{17}$
        cm$^{-2}$/kms$^{-1}$, n(H$_{2}$)= 10$^{4}$-10$^{5}$ cm$^{-3}$
        and $T_{K}=$100 K. We suggest a higher value for the gas
        density and lower values for N($^{12}$CO)/$\Delta v=3\times
        10^{16}$cm$^{-2}$/kms$^{-1}$ and $T_{K}= 25$K than those
        proposed by \citet{Peti98a}. The differences can be explained
        by the fact that we do not use the same set of CO line
        intensity ratios : \citet{Peti98a} used line ratios combining
        low-J CO lines ($^{12}$CO(2-1), $^{13}$CO(2-1), $^{12}$CO(3-2)
        and $^{13}$CO(3-2)) while we used a larger number of intensity
        ratios focussed on the high-J CO lines (See Table
        ~\ref{tab:raplvg}). For IC 342, \citet{Isra03} fit their data
        using a LVG model with N($^{12}$CO)/$\Delta
        v=$6-10$\times10^{16}$ cm$^{-2}$/kms$^{-1}$, T$_{K}=$100-150 K
        and n(H$_{2}$)= 3.0$\times 10^{3}$ cm$^{-3}$. \citet{Ecka90c}
        suggested a model with N($^{12}$CO)=3-4$\times 10^{18}$
        cm$^{-2}$, T$_{K}>$20 K and a density around
        n(H$_{2}$)$\approx$ 2.0$\times 10^{3}$ cm$^{-3}$ for the
        center of IC 342. \citet{Meie00} deduced from their
        observations a beam-averaged density of n(H$_{2}$)= 1.3$\times
        10^{3}$ cm$^{-3}$. Our model agrees very well with the gas
        density values proposed in the two latter articles. The fitted
        N($^{12}$CO)/$\Delta v$ value is intermediate between results
        of \citet{Isra03} and \citet{Ecka90c}. For $T_{K}$, we agree
        better with the value proposed by \citet{Ecka90c} than with
        the one proposed by \citet{Isra03}. For NGC 4038,
        \citet{Zhu03} proposed a one component model
        (N($^{12}$CO)/$\Delta v=$3.4$\times 10^{16}$
        cm$^{-2}$/kms$^{-1}$, n(H$_{2}$)= 1-4$\times 10^{3}$ cm$^{-3}$
        and $T_{K}=$43 K) in good agreement with our study except for
        the n(H$_{2}$) value. For Overlap, the one component model
        proposed by \citet{Zhu03} has N($^{12}$CO)/$\Delta
        v\approx$1.0$\times 10^{16}$ cm$^{-2}$/kms$^{-1}$, n(H$_{2}$)=
        1-5$\times 10^{3}$ cm$^{-3}$ and $T_{K}=$33 K for their
        (0'';10'') offset positions which almost corresponds to our
        Overlap position. The agreement of the former study with our
        values is not as good as for NGC 4038 but \citet{Zhu03} used a
        higher CO abundance ratio ($ ^{12}$CO/$ ^{13}$CO = 60) than
        we do (40), which may be at the origin of the differences. For M 83,
        \citet{Isra01} presented two LVG models: one with
        N($^{12}$CO)/$\Delta v=$1-3$\times 10^{17}$
        cm$^{-2}$/kms$^{-1}$, $T_{K}=$30-150 K and
        n(H$_{2}$)=0.5-3.0$\times 10^{3}$ cm$^{-3}$, and one with
        N($^{12}$CO)/$\Delta v=$0.06-1 $\times 10^{17}$
        cm$^{-2}$/kms$^{-1}$, $T_{K}=$60-100 K and n(H$_{2}$)=
        0.03-1.0$\times 10^{5}$ cm$^{-3}$. Our fitted parameters compare
        well with both models, although they are closer to
        the second model conditions. For NGC 6946, \citet{Wals02}
        proposed the following LVG model: N($^{12}$CO)/$\Delta
        v=$2.9$\times 10^{16} $ cm$^{-2}$/kms$^{-1}$, $T_{K}=$40 K and
        n(H$_{2}$)= 2.0$\times 10^{3}$ cm$^{-3}$. \citet{Isra01}
        presented two other possible LVG models: one with
        N($^{12}$CO)/$\Delta v= 1 - 10 \times 10^{17} $
        cm$^{-2}$/kms$^{-1}$, $T_{K}= 30-150$ K and n(H$_{2}$)= $0.5 -
        1.0 \times 10^{3}$ cm$^{-3}$, and one with
        N($^{12}$CO)/$\Delta v= 3-6 \times 10^{16}$
        cm$^{-2}$/kms$^{-1}$, $T_{K}= 30-150$ K and n(H$_{2}$)=
        $0.1-1.0 \times 10^{4}$ cm$^{-3}$. All these models are consistent
        with the parameters obtained in the present work.

    \subsection{PDR models}\label{secsub:pdr}

        \subsubsection{Fitting procedures}\label{secsubsub:pdrfit}

        We use the PDR models developed by Le Bourlot et al. (1993) for
        Galactic sources (see also \citealt{LePe02}). The source is modelled
        as a plane-parallel slab, illuminated on both sides by FUV radiation
        to better reproduce a starburst environment where massive stars, and
        giant molecular clouds, are spatially correlated. Model parameters
        include the gas density, assumed uniform, the intensity of the
        illuminating FUV radiation, the gas phase elemental abundances, the
        grain properties and the gas to dust ratio.

        We considered that all galaxies have a metallicity close to solar (See
        Table ~\ref{tab:prop}), and we used Milky Way abundances in our models
        ($12+log(\frac{O}{H})=$8.90 $\pm$ 0.04 \citealt{Arim96}). For IC 10 and
        IC 342, this hypothesis may be rather crude considering the lower and
        higher (respectively) metallicity of their nuclei
        ($12+log(\frac{O}{H})= 8.31\pm 0.2$ in \citealt{Arim96} and
        $12+log(\frac{O}{H})=$9.30 in \citealt{Vila92}). Therefore results
        obtained on these sources should be considered more
        cautiously. Standard grain properties and a gas to dust ratio
        appropriate for Galactic interstellar clouds have been
        adopted. The $^{12}$C/$^{13}$C ratios are the same as used
        above for the LVG models (see Table ~\ref{tab:resmod}). We
        have sampled a wide range of parameter space, varying the gas
        density, n(H) from 1.0$\times 10^{2}$ cm$^{-3}$ to 1.0$\times
        10^{7}$ cm$^{-3}$, and the incident FUV flux, $\chi_{FUV}$,
        from $1.5 \times 10^{3}$ G$_{0}$ to $1.5 \times 10^{6}$
        G$_{0}$, where G$_{0}$ is the local average interstellar
        radiation field (ISRF) determined by Draine (1978) (G$_{0}=
        2.7\times 10^{-3}$ $ergcm^{-2}s^{-1}$). We fit the PDR models
        with the same line intensity ratios   used to fit the the
        LVG models (See Table
        ~\ref{tab:rappdr}) but computed from values
        listed in Table ~\ref{tab:obs2} (with an asterisk) expressed
        in Wm$^{-2}$sr$^{-1}$, a unit more appropriate for PDR model
        calculations. Indeed, the output values (emissivities) from
        PDR models calculations are expressed in
        ergcm$^{-2}$s$^{-1}$sr$^{-1}$.

        The best model is obtained using a least square fitting method as
        applied in Sect. ~\ref{secsub:lvg} taking into account the errors
        on the observations.

        \subsubsection{Results}\label{secsubsub:pdrres}

        Best fit solutions (physical
        parameters) for the selected galaxies are presented in
        Table ~\ref{tab:resmod}. Model predictions for the CO line
        emissivities are shown in the right side of Figs.~\ref{fig:pred_ic10},
        ~\ref{fig:pred_ic342}, ~\ref{fig:pred_n4038}, ~\ref{fig:pred_Overlap},
        ~\ref{fig:pred_m83} and ~\ref{fig:pred_n6946} for IC 10, IC 342,
        NGC 4038, Overlap, M 83 and NGC 6946, respectively. Model
        predictions have been scaled by a surface filling factor
        (FF$_{PDR}$ listed in Table ~\ref{tab:resmod})
        to match the observed line intensities.
        As stated above, PDR models have been developed for local
        interstellar clouds. The velocity dispersion of the modelled cloud is
        a parameter in those models, which is used for computing the
        photo-dissociation rates of H$_{2}$ and CO, and the line
        emissivities. This parameter is set to 1 kms$^{-1}$, a typical figure
        for local molecular clouds (see \citealt{Wolf90}). Using a PDR model
        for fitting galaxy observations is complicated by the fact that, in a
        galaxy, many PDRs contribute to the signal detected in each beam,
        resulting in a broad line (tens to hundreds of kms$^{-1}$), compared
        to a single PDR line (1 kms$^{-1}$). To correct for this effect, the
        modelled emissivities have been multiplied by the line width ratio,
        $\frac{\Delta v(galaxy)}{\Delta v(PDR)}$, where $\Delta
        v(galaxy)$ is the line width of the CO lines as reported
        in Table ~\ref{tab:resmod}, and where $\Delta v(PDR)=$1
        kms$^{-1}$. Once this correction is performed, PDR model
        results are compared with observed data in the same way as the
        LVG models (see Sect.~\ref{secsub:lvg}). The surface
        filling factors of the emission in the beam, FF$_{PDR}$, are also
        computed (and listed in Table ~\ref{tab:resmod}) and they are in
        reasonable agreement with those  derived from LVG models
        for all sources.

        The PDR solutions obtained from the least square fitting procedure
        are described below for the individual galaxies. The $^{12}$CO and
        the $^{13}$CO cooling rates are summarized in Table ~\ref{tab:resmod}.

        \begin{itemize}
            \item [*] For IC 10, the ``best'' model is represented in
            Fig. ~\ref{fig:pred_ic10} with grey squares. The
            $^{12}$CO(5-4) line represents 24.7\% of the total
            $^{12}$CO cooling rate while the $^{12}$CO(6-5) and
            $^{12}$CO(4-3) lines correspond to 24.6\% and 16.8\%,
            respectively. The main lines for the $^{13}$CO cooling
            rate are the $^{13}$CO(5-4) (33.0\%), $^{13}$CO(6-5)
            (26.8\%) and $^{13}$CO(4-3) (25.8\%) lines.\\

            \item [*] For IC 342, the ``best'' model is represented in
            Fig. ~\ref{fig:pred_ic342} with grey squares. The lines
            which contribute the most to the $^{12}$CO cooling rate
            are the $^{12}$CO(5-4) (27.3\% of the total intensity),
            $^{12}$CO(6-5) (23.6\%) and $^{12}$CO(4-3) (20.1\%)
            lines. The lines which contribute the most to the
            $^{13}$CO cooling rate are the $^{13}$CO(4-3) (30.6\%),
            $^{13}$CO(5-4) (29.6\%) and $^{13}$CO(3-2) (17.7\%)
            lines.\\

            \item [*] For NGC 4038, the ``best'' model is represented in
            Fig. ~\ref{fig:pred_n4038} with grey squares. The lines which
            contribute the most to the $^{12}$CO and to the $^{13}$CO cooling
            rates are the $^{12}$CO(6-5) (23.4\% of the total intensity),
            $^{12}$CO(5-4) (20.3\%), $^{12}$CO(7-6) (19.0\%) lines
            and the $^{13}$CO(6-5) (35.7\%), $^{13}$CO(5-4)
            (33.4\%) and $^{13}$CO(4-3) (21.1\%) lines, respectively.\\

            \item [*] For the Overlap region, the ``best'' model is
            represented in Fig. ~\ref{fig:pred_Overlap} with grey
            squares. The lines which contribute the most to the
            $^{12}$CO and to the $^{13}$CO cooling rates are the
            $^{12}$CO(6-5) (22.9\% of the total intensity),
            $^{12}$CO(7-6) (19.7\%), $^{12}$CO(5-4) (19.1\%) lines and
            the $^{13}$CO(6-5) (34.7\%), $^{13}$CO(5-4) (33.3\%) and
            $^{13}$CO(4-3) (21.6\%) lines, respectively.\\

            \item [*] For M 83, the ``best'' model is represented in
            Fig. ~\ref{fig:pred_m83} with grey squares. The lines which
            contribute the most to the $^{12}$CO cooling rate are the
            $^{12}$CO(6-5) (20.4\% of the total intensity),
            $^{12}$CO(7-6) (20.4\%) and $^{12}$CO(5-4) (15.4\%) lines.
            The lines which contribute the most to the $^{13}$CO cooling rate
            are the $^{13}$CO(6-5) (34.1\%), $^{13}$CO(5-4) (32.9\%)
            and $^{13}$CO(4-3) (21.9\%) lines. \\

            \item [*] For NGC 6946, the ``best'' model is represented
            in Fig. ~\ref{fig:pred_n6946} with grey squares. The lines
            which contribute the most to the $^{12}$CO cooling rate
            are the $^{12}$CO(5-4) (28.5\% of the total intensity),
            $^{12}$CO(4-3) (22.7\%) and $^{12}$CO(6-5) (21.6\%)
            lines. The main lines for the $^{13}$CO cooling rate are
            the $^{13}$CO(5-4) (33.0\%), $^{13}$CO(4-3) (27.8\%) and
            $^{13}$CO(6-5) (23.7\%) lines.\\
        \end{itemize}

        For the IC 10 nucleus (Fig.~\ref{fig:pred_ic10}), we obtained
        a good agreement between the two solutions proposed by the LVG
        and PDR models (similarly shaped curves and $^{12}$CO cooling
        rates); probably because of the high data quality. Small
        differences appear in the $^{13}$CO cooling curves since the
        peak seems to be located at the $^{13}$CO(4-3) transition for
        the LVG model while the PDR peak clearly corresponds to the
        $^{13}$CO(5-4) transition. $^{13}$CO(6-5) observations would
        be very useful to discriminate between the model solutions.

        In IC 342 (Fig.~\ref{fig:pred_ic342}), we observed an
        excellent agreement between the two model solutions, the sole
        difference being in the peak position for the $^{13}$CO
        cooling curves (the LVG peak corresponding to the
        $^{13}$CO(3-2) line while the PDR one being located at
        $^{13}$CO(4-3)). Here also, $^{13}$CO(6-5) data would be very
        useful to better localize the $^{13}$CO peak. In this case, we
        have a very complete and high quality dataset. The $^{12}$CO
        cooling rate deduced from the PDR model is very close to the
        $^{12}$CO cooling rate computed from the LVG model.

        At the Antennae positions (Overlap and NGC 4038 in
        Figs. ~\ref{fig:pred_n4038} and ~\ref{fig:pred_Overlap}), LVG
        models, as well as PDR models, do not reproduce observations
        adequately due to the low signal-to-noise ratio of both
        $^{12}$CO(7-6) and $^{12}$CO(4-3) lines. Indeed, because these
        latter observations are weaker than we expected, we have not
        considered them in the least square fitting
        procedures. However, for NGC 4038, both model solutions (LVG
        and PDR) are coherent : the maximum of the $^{12}$CO cooling
        curve deduced from the LVG model is the same as the result
        from the PDR model ($^{12}$CO(6-5)) and the $^{12}$CO cooling
        rates are similar. For Overlap, the maxima of the LVG and PDR
        $^{12}$CO cooling curves appear both for $^{12}$CO(6-5) and
        the predicted $^{12}$CO cooling rates from the LVG and PDR
        models are close to each other. Despite this, we clearly see
        differences in the predicted line intensities from LVG and PDR
        models, certainly due to the lack of relevant $^{12}$CO(7-6)
        detections. Concerning the $^{13}$CO cooling curves shown in
        the right side of Fig. ~\ref{fig:pred_n4038} (for NGC 4038)
        the two model solutions are compatible. Consequently, the
        total $^{13}$CO cooling rate agrees between the LVG and PDR
        models (see Table ~\ref{tab:resmod}). For the Overlap position
        (Fig. ~\ref{fig:pred_Overlap}), the LVG and PDR $^{13}$CO
        cooling curves peak at the different positions ($^{13}$CO(6-5)
        for the PDR and $^{13}$CO(5-4) for the LVG) but the $^{13}$CO
        cooling rate are different by a factor $\lesssim$ 3 (see Table
        ~\ref{tab:resmod}). For NGC 4038 and Overlap, $^{13}$CO(6-5)
        observations would be very useful to discriminate between
        model solutions.

        For M 83, the lack of $^{12}$CO(7-6) data explains the
        differences (noticeable in Figs.~\ref{fig:pred_m83}) between
        the LVG and PDR solutions for the $^{12}$CO cooling
        curves. The peak corresponds to the $^{12}$CO(7-6) line in the
        LVG and PDR models. Observations of the $^{12}$CO(7-6),
        $^{12}$CO(8-7) lines and possibly up would be very
        useful. Despite the different shape of the $^{12}$CO cooling
        curves, very similar $^{13}$CO cooling curves are obtained. We
        notice that although the M 83 dataset is less complete than
        for other sources, we obtained fairly similar values of the
        LVG and PDR $^{12}$CO cooling rates (see Table ~\ref{tab:resmod}).

        For NGC 6946, the peak positions of the $^{12}$CO cooling curves are
        similar ($^{12}$CO(5-4)) for the LVG and the PDR models. Also in
        this case, observations of the $^{12}$CO(7-6) and $^{12}$CO(8-7) lines
        appear essential to discriminate between model solutions.
        For the $^{13}$CO cooling curve, we observed stronger
        differences especially for the peak position since the LVG maximum
        corresponds to the $^{13}$CO(3-2) line while the PDR maximum
        appears for the $^{13}$CO(5-4) line. Once again, the $^{13}$CO(6-5)
        observation would be very useful to better localize the $^{13}$CO peak.

        \begin{table*}
            \caption{Observed and predicted (``best'' PDR model) line
            intensity ratios.}\label{tab:rappdr}
        \begin{center}
                \begin{tabular}{|c|c|c|c|c|c|}
                \hline
                I & IC 10 & IC 342 & I & NGC 4038 & OVERLAP\\
                (Wm$^{-2}$sr$^{-1}$)& obs.$^{*}$ & obs.$^{*}$ &
                (Wm$^{-2}$sr$^{-1}$) & obs.$^{*}$ & obs.$^{*}$\\
                \hline
                $\frac{^{12}CO(3-2)}{^{12}CO(4-3)}$ &
                0.8$\pm$0.1$^{a}$ & 0.9$\pm$0.1$^{a}$ &
                $\frac{^{12}CO(3-2)}{^{12}CO(4-3)}$ &
                0.6$\pm$7.7$\times 10^{-2}$ & 1.0$\pm$0.1\\
                \hline
                $\frac{^{12}CO(3-2)}{^{12}CO(6-5)}$ &
                0.3$\pm$6.3$\times 10^{-2,a}$ & 0.5$\pm$4.5$\times
                10^{-2,a}$ & $\frac{^{12}CO(3-2)}{^{12}CO(6-5)}$ &
                0.1$\pm$2.4$\times 10^{-2,a}$ & 0.2$\pm$9.4$\times
                10^{-2,a}$\\
                \hline
                $\frac{^{12}CO(3-2)}{^{12}CO(7-6)}$ &
                0.7$\pm$0.2$^{a}$ & 1.2$\pm$0.1$^{a}$ &
                $\frac{^{12}CO(3-2)}{^{12}CO(7-6)}$ &
                1.8$\pm$0.6$^{a}$ & 1.2$\pm$0.2$^{a}$\\
                \hline
                $\frac{^{12}CO(2-1)}{^{12}CO(7-6)}$ &
                0.3$\pm$0.1$^{a}$ & 0.4$\pm$0.1
                &$\frac{^{12}CO(2-1)}{^{12}CO(6-5)}$ &
                (5.8$\pm$1.9)$\times 10^{-2,a}$ & (8.3$\pm$4.9)$\times
                10^{-2,a}$\\
                \hline
                \hline
                $\frac{^{12}CO(3-2)}{^{13}CO(3-2)}$ &
                11.8$\pm$2.3$^{a}$ & 6.3$\pm$0.8$^{a}$ &
                $\frac{^{12}CO(3-2)}{^{13}CO(3-2)}$ &
                20.6$\pm$12.6$^{a}$ & 34.5$\pm$9.4$^{a,b}$\\
                \hline
                $\frac{^{12}CO(1-0)}{^{13}CO(1-0)}$ & - & 12.9$\pm$0.5 &
                $\frac{^{12}CO(1-0)}{^{13}CO(1-0)}$ & - & -\\
                \hline
                $\frac{^{12}CO(2-1)}{^{13}CO(2-1)}$ & 17.6$\pm$7.0 &
                12.5$\pm$3.0 & $\frac{^{12}CO(2-1)}{^{13}CO(2-1)}$ &
                27.9$\pm$11.1 & 18.6$\pm$7.5\\
                \hline
                & PDR model & PDR model & & PDR model & PDR model\\
                \hline
                $\frac{^{12}CO(3-2)}{^{12}CO(4-3)}$ & 0.5 & 0.5 &
                $\frac{^{12}CO(3-2)}{^{12}CO(4-3)}$ & 0.4 & 0.4\\
                \hline
                $\frac{^{12}CO(3-2)}{^{12}CO(6-5)}$ & 0.3 & 0.4 &
                $\frac{^{12}CO(3-2)}{^{12}CO(6-5)}$ & 0.2 & 0.2\\
                \hline
                $\frac{^{12}CO(3-2)}{^{12}CO(7-6)}$ & 0.5 & 0.9 &
                $\frac{^{12}CO(3-2)}{^{12}CO(7-6)}$ & 0.3 & 0.2\\
                \hline
                $\frac{^{12}CO(2-1)}{^{12}CO(7-6)}$ & 0.1 & 0.3 &
                $\frac{^{12}CO(2-1)}{^{12}CO(6-5)}$ & 4.5$\times
                10^{-2}$ & 4.9$\times 10^{-2}$\\
                \hline
                \hline
                $\frac{^{12}CO(3-2)}{^{13}CO(3-2)}$ & 12.3 & 6.3 &
                $\frac{^{12}CO(3-2)}{^{13}CO(3-2)}$ & 19.6 & 15.7 \\
                \hline
                $\frac{^{12}CO(1-0)}{^{13}CO(1-0)}$ & 18.9 & 11.3 &
                $\frac{^{12}CO(1-0)}{^{13}CO(1-0)}$ & 24.6 & 22.6 \\
                \hline
                $\frac{^{12}CO(2-1)}{^{13}CO(2-1)}$ & 14.6 & 7.1 &
                $\frac{^{12}CO(2-1)}{^{13}CO(2-1)}$ & 21.9 & 18.5 \\
                \hline
                \end{tabular}
            \end{center}
            \vspace*{-0.4cm}
            \hspace*{2.05cm}
                \begin{tabular}{|c|c|c|}
                \hline
                & M 83 & NGC 6946 \\
                & obs.$^{*}$ & obs.$^{*}$ \\
                \hline
                $\frac{^{12}CO(3-2)}{^{12}CO(4-3)}$ &
                0.5$\pm$0.2$^{a}$ & 0.6$\pm$0.1$^{a}$\\
                \hline
                $\frac{^{12}CO(3-2)}{^{12}CO(6-5)}$ &
                0.1$\pm$3.9$\times 10^{-2,a}$ & 0.6$\pm$0.1$^{a}$\\
                \hline
                $\frac{^{12}CO(3-2)}{^{12}CO(7-6)}$ & - & - \\
                \hline
                $\frac{^{12}CO(2-1)}{^{12}CO(6-5)}$ &
                (5.2$\pm$3.8)$\times 10^{-2,a}$ & 0.3$\pm$5.9$\times
                10^{-2,a}$\\
                \hline
                \hline
                $\frac{^{12}CO(3-2)}{^{13}CO(3-2)}$ &
                32.7$\pm$10.1$^{a}$ & 19.5$\pm$2.3$^{a}$\\
                \hline
                $\frac{^{12}CO(1-0)}{^{13}CO(1-0)}$ & - & 12.7$\pm$5.1\\
                \hline
                $\frac{^{12}CO(2-1)}{^{13}CO(2-1)}$ &
                5.1$\pm$0.6$\times 10^{-2}$ & 34.0$\pm$7.6\\
                \hline
                & PDR model & PDR model\\
                \hline
                $\frac{^{12}CO(3-2)}{^{12}CO(4-3)}$ & 0.4 & 0.5\\
                \hline
                $\frac{^{12}CO(3-2)}{^{12}CO(6-5)}$ & 0.2 & 0.5\\
                \hline
                $\frac{^{12}CO(3-2)}{^{12}CO(7-6)}$ & 0.2 & 1.0\\
                \hline
                $\frac{^{12}CO(2-1)}{^{12}CO(6-5)}$ & 5.1$\times
                10^{-2}$ & 0.1\\
                \hline
                \hline
                $\frac{^{12}CO(3-2)}{^{13}CO(3-2)}$ & 9.8 & 18.3\\
                \hline
                $\frac{^{12}CO(1-0)}{^{13}CO(1-0)}$ & 19.3 & 22.2\\
                \hline
                $\frac{^{12}CO(2-1)}{^{13}CO(2-1)}$ & 12.6 & 20.1\\
                \hline
                \end{tabular}

                $^{*}$ : ratio derived from observations marked with
                asterisks in Table ~\ref{tab:obs2}; $^{a}$ : values
                used as constraints for the LVG models; $^{b}$ : value
                from Table 2 in \citet{Zhu03}; - : values not observed
                neither found in literature.\\
        \end{table*}

        \subsubsection{Discussion}\label{secsubsub:pdrdis}

        As PDR models provide predictions for lines of C$^{+}$, O, and
        C together with CO rotational lines, it is interesting to
        check whether the proposed PDR solutions are appropriate to
        reproduce observation of other cooling lines such as the fine
        structure lines of atomic carbon. In fact, for all sources
        studied here (see Table ~\ref{tab:resmod}), the predicted C
        cooling rates are typically larger than those observed one by
        significant factors (between 4.5 and 50).

        For a large majority of the sample galaxies, we conclude that
        PDR models focussed on the high-J CO lines do not correctly reproduce
        the observed atomic carbon data well. To better model the observed
        atomic carbon transitions, another set of physical conditions
        might be a better choice. For example, we could have increased
        the total extinction A$_{v}$, set to 10 mag in the PDR models
        we used. It is expected that the relative contributions of C
        and CO will change as CO lines are produced at larger depths
        than C. We performed a few calculations, variyng the total
        extinction across the slab, A$_{v}$. As expected, Co lines
        become more prominent relative to C lines with increased
        A$_{v}$. However, the observed ratio is never reached. A more
        completed study of the parameter space is needed, which is
        beyond the scope of this paper.


\section{Comparison with the Center of Milky Way and the Cloverleaf
  QSO}\label{sec:Comp}

\begin{figure}
  \begin{center}
    \epsfxsize=9cm
    \epsfbox{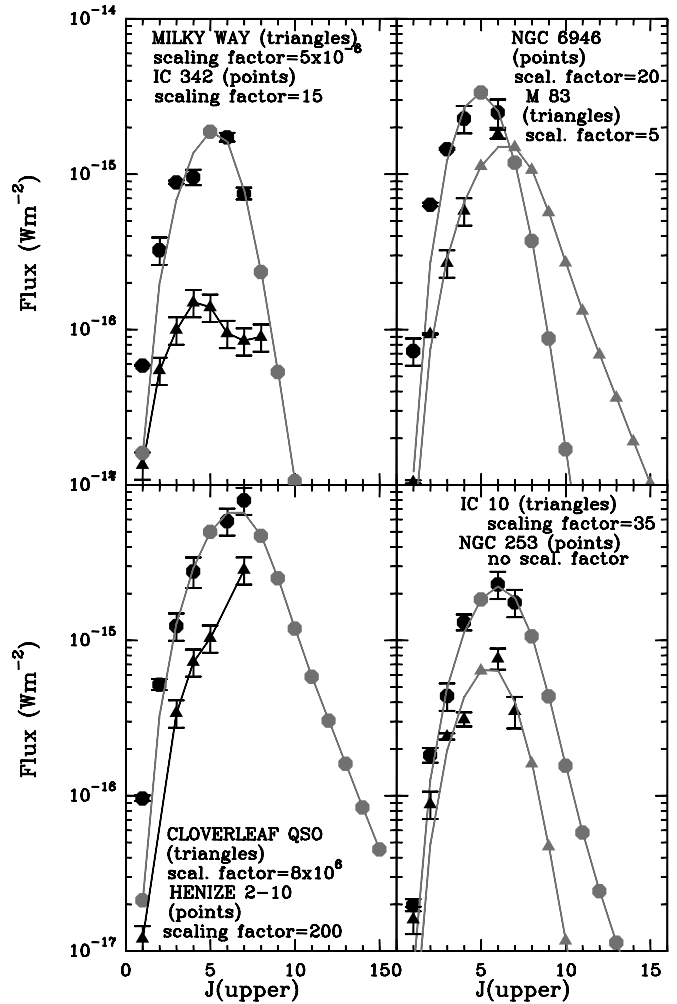}
    \caption{Flux (Wm$^{-2}$) vs. $J_{upper}$ for the center of the
    Milky Way (top left, triangles), IC 342 (top left, points), NGC
    6946 (top right, points), M 83 (top right, triangles), the
    Cloverleaf QSO (bottom left, triangles), Henize 2-10 (bottom
    left, points) IC 10 (bottom right, triangles) and NGC 253 (bottom
    right, points). PDR models (grey points or grey triangles) have
    been used to obtained line fluxes of unobserved lines (eg. $^{12}$CO(5-4),
    $^{12}$CO(7-6), $^{12}$CO(8-7), $^{12}$CO(9-8),...) depending
    on the completeness of the dataset obtained for each source.
    Observations are shown in black with error bars.
    To ease the comparison, we applied scaling factors on all sources
    data except for NGC 253. The value of these scaling factors is
    specified for each source on the plots. We notice that for the
    Cloverleaf QSO, the $^{12}$CO(1-0) line flux is an upper limit (see
    \citealt{Tsub99}).}\label{fig:comp}
  \end{center}
\end{figure}

It is interesting to compare this galaxy sample with other, well
known, sources. As we previously discussed previously, the peak
position of the $^{12}$CO cooling curves shifts between nuclei. In
Fig. ~\ref{fig:comp} and ~\ref{fig:comp2}, we compare the $^{12}$CO
cooling curves for the center of the Milky Way and the Cloverleaf QSO
[HB89] 1413+117 with our sample galaxies. We completed the set of
observed $^{12}$CO lines with predictions from PDR models up to
$^{12}$CO(15-14). Because the PDR model for Overlap is less reliable,
we excluded this source from this study. We have kept NGC 4038
because LVG and PDR model results are consistent for this object.
Flux values for the center of the Milky Way and for the Cloverleaf
QSO are from \citet{Fixs99} and \citet{Barv97, Tsub99, Weis03},
respectively.

The CO cooling rates (listed in Table ~\ref{tab:resmod})
appear to be higher in the starburst environments (NGC 253 in
\citealt{Baye04}, M 83) than in the normal spiral galaxy (IC 342, NGC
6946) or irregular galaxies (Henize 2-10 in \citealt{Baye04} and IC
10). This phenomenon seems not to be solely a distance effect since we
obtained the highest observed CO cooling rates for NGC 253 and M 83,
two galaxies which are neither the nearest nor the farthest sources
(2.5 Mpc and 3.5 Mpc, respectively). Moreover, the lowest observed CO
cooling rates are obtained for Henize 2-10 and IC 10 which distances
are 6 Mpc and 1 Mpc, respectively.

By looking at Fig. ~\ref{fig:comp} and  ~\ref{fig:comp2}, we can
distinguish three different behaviors : for Henize 2-10 and the
Cloverleaf QSO, the CO cooling curve peaks at the $^{12}$CO(7-6) line
or up;  for IC 10, NGC 253, NGC 4038, M 83 and NGC 6946, the turnover
appears at the $^{12}$CO(6-5) line\footnote{For NGC 6946, due to the
lack of the $^{12}$CO(7-6) detection, the LVG and PDR predictions are
less reliable than for the other sources for which we have a more
complete dataset. To determine the position of the NGC 6946 CO
cooling curve, we chose to trust the observations rather than the
predictions.}; and for the center of the Milky Way and IC 342, the
turnover is found near the $^{12}$CO(4-3) line. We may ask whether
these observed differences are solely consequences of the difference
in linear resolution solely, or whether they are due to differences
in physical conditions.

The turnover of the $^{12}$CO cooling curve does not depend on the
distance, since there is no obvious correlation between the distance
and the position of the CO peak. For instance, the fact that the
Cloverleaf QSO and Henize 2-10 have very similar CO cooling curve
shapes (See Fig. ~\ref{fig:comp}), is related to the close
similarities of the physical properties of the warm gas for these two
sources, which translates into similar CO line ratios.

\begin{figure}
  \begin{center}
    \epsfxsize=5cm
    \epsfbox{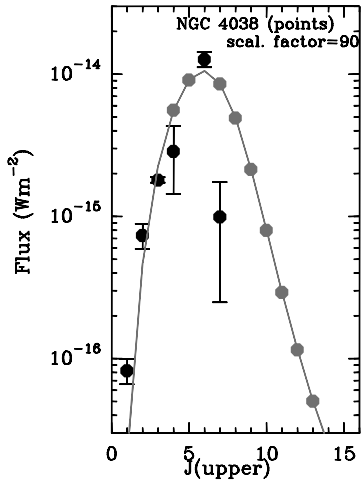}
    \caption{Flux (Wm$^{-2}$) vs. $J_{upper}$ for the
    NGC 4038 nucleus (points). See the caption of Fig.~\ref{fig:comp}.
    In this figure, we plot observations of the $^{12}$CO(4-3) and
    $^{12}$CO(7-6) lines even though they have a low signal-to-noise
    ratio. We also plot their PDR predicted fluxes.}\label{fig:comp2}
  \end{center}
\end{figure}

The position of the peak in the CO cooling curves is a particularly
interesting parameter, as it moves towards higher J for more actively
star forming galaxies. Indeed, such curves could be used to diagnose
the gas properties and measure the cooling rate in galaxies. Few
detections are necessary (high-J CO lines as $^{12}$CO(6-5) and
$^{12}$CO(7-6)) to obtain a first estimation of the gas cooling rate.


\section{Conclusions}

In this paper, we have presented observations of the C and
CO submillimeter lines (up to the $^{12}$CO(7-6) line at 806 GHz and the
[CI]($^{3}$P$_{2}$-$^{3}$P$_{1}$) line at 809 GHz) for the Antennae
galaxies (NGC 4038 and the overlap region between the two nuclei), the
nuclei of IC 10, IC 342, M 83 and NGC 6946 (see Tables
~\ref{tab:obs1} and ~\ref{tab:obs2}). We also detected submillimeter
C and CO lines for the following galaxies: nucleus of Arp 220, Centaurus A,
IRAS 10565+2448, M 51, M 82, Markarian 231, NGC 3079, NGC 4736, NGC
6090 nuclei and in the spiral arms of M 83 and NGC 6946.

We succeed in fitting all observed CO lines to accurately estimate the
CO cooling rate. LVG and PDR models have been used for fitting the CO
line intensities. As we selected
sources with different morphological types, we have compared the
contribution of C and CO and the shape of the CO cooling curves
between galaxies.

\begin{itemize}
    \item [1] The total CO cooling rates were estimated summing
    intensities (I in Wm$^{-2}$sr$^{-1}$) of all CO transitions up to
    $^{12}$CO(15-14), derived either from observations or from PDR
    model predictions. We showed that the total CO cooling rate is
    higher in the starburst nuclei (M 83 and NGC 253) than in
    normal galaxies (IC 342 and NGC 6946) and irregular
    galaxies (IC 10 and Henize 2-10).\\

    \item [2] The shape of the CO cooling curve depends on the galaxy
    activity. The CO cooling is maximum for the the $^{12}$CO(7-6) line
    and up for Henize 2-10 and the Cloverleaf QSO, while for IC
    10, NGC 253, NGC 4038, M 83 and NGC 6946 the turnover appears at
    the $^{12}$CO(6-5) line. For the center of the Milky Way and IC
    342, the turnover is located at $^{12}$CO(4-3) line. These
    lines are the main CO cooling lines, which shows
    the predominant role of the high-J CO lines ($\gtrsim
    ^{12}$CO(5-4)) in the molecular
    gas cooling. {\em The shape of the
    CO cooling curve can be used as a diagnostic of the galaxy star
    forming activity}.\\

    \item [3] Whatever the morphological type, the C cooling rates are
    lower than the CO cooling rates (the ratio varies from a factor of
    $\lesssim 4.0$ for IC 10 and IC 342, of $\geqslant$ 10 for NGC 253
    and Henize 2-10 (reported in \citealt{Baye04}). The factor are
    estimated to be $\approx$ 20 for NGC 4038, Overlap,
    M 83 and NGC 6946. As we missed the
    [CI]($^{3}$P$_{2}$-$^{3}$P$_{1}$) line at 809 GHz for the Antennae, M
    83 and NGC 6946, the latter figures should be confirmed
    with more observations.\\
\end{itemize}

Our analysis made use of two complementary models: LVG models (Large
Velocity Gradient for radiative transfer calculations) and PDR
models (combining chemistry, thermal balance and radiative transfer)
for deriving the physical conditions in the galaxy nuclei, and
for computing the intensity of the missing CO lines.
 We obtained the following results:

\begin{itemize}
    \item [4] We succeeded in deriving physical properties (the
    kinetic
    temperature (T$_{K}$), the gas density (n(H$_{2}$)), the CO column
    density divided by the line width (N($^{12}$CO)/$\Delta v$) and the Far-UV
    radiation field ($\chi_{FUV}$)) of the warm and dense molecular gas,
    using high-J CO and $^{13}$CO line intensity ratios for
    constraining the models. The LVG models alone can be highly
    degenerated
    as an extended region in the
    (n(H$_{2}$); $T_{K}$) parameter space provides equally good fits. However,
    best models provide consistent estimations of the CO cooling rates.\\

    \item [5] The surface filling factor of the molecular emission
    amounts to 1-7\% in a 21.9''beam. High resolution
    observations, using existing (IRAM-PdBI, SMA) or planned (CARMA,
    ALMA) (sub)millimeter interferometers, should be able to resolve
    individual sources in these galaxies. \\

    \item [6] The $^{12}$CO(5-4) line appears as one of the main CO
    cooling line for most sources. Unfortunately, the line is blocked
    by the atmosphere for local galaxies. Spatial missions (eg the
    Herschel Space Observatory) will soon provide measurements of this
    important line. We also
    showed the key role of the $^{13}$CO(6-5) and $^{12}$CO(8-7)
    lines as constraints to PDR and LVG models. Future observations with
    ALMA and the Herschel satellite will provide
    these informations.\\

    \item [7] While $^{12}$CO(3-2) and $^{12}$CO(4-3) lines are detected in
    star forming regions outside galaxy nuclei, we could not detect
    $^{12}$CO(6-5) emission there because of the lack of sensitivity. A similar
    analysis on spiral arm regions should be performed in the near
    future.\\

    \item [8] For our galaxy nuclei sample, PDR solutions fitted to
    the CO lines, do not provide a good match to the fine structure
    line intensities of neutral carbon.\\

\end{itemize}


\begin{acknowledgements}
    This work has benefitted from financial support from the CNRS/INSU
    research programs PCMI \& PNG.
    We thank J. Cernicharo and M. Perault for letting us use their CO LVG
    models and J. Le Bourlot for introducing us to the
    Meudon PDR model. We made use of the SIMBAD and NED data base for helping
    us finding basic galaxy properties. The CSO is funded by the NSF under contract \# AST
    9980846.
\end{acknowledgements}


\bibliographystyle{aa}
\bibliography{AA_2005_3872}

%
\appendix

\section{Convolution formulae}\label{cal}
As CO lines have been observed with different spatial resolutions,
the line intensities should be convolved to the same linear
resolution before performing meaningful comparisons. The final
resolution of 21.9'' has been chosen since it corresponds
to the beaim size of the CSO when observing at 345 GHz (frequency of the
$^{12}$CO(3-2)). To perform this convolution, we modelled emitting
region of each galaxy using Gaussian profiles either axisymmetric
(See Eq. ~\ref{source-ronde}) or elliptical (See Eq.
~\ref{source-ellipt}) :

\begin{center}
    \begin{equation}\label{source-ronde}
        f_{s}(x,y) = \text{f}_{0}\frac{1}{\pi \sigma_{s}^{2}}
        \exp{-\frac{x^{2}+y^{2}}{\sigma_{s}^{2}}}
    \end{equation}
\end{center}

\begin{center}
    \begin{equation}\label{source-ellipt}
        f_{s}(x,y) = \text{f}_{0}\frac{1}{\pi \sigma_{s,x} \sigma_{s,y}}
        \exp{-\frac{x^{2}}{\sigma_{s,x}^{2}}}
        \exp{-\frac{y^{2}}{\sigma_{s,x}^{2}}}
    \end{equation}
\end{center}

where f$_{0}$ is a normalisation constant and where the Full Width
Half Maximum (FWHM) is FWHM$_{s}$ = 2$\sigma_{s}\sqrt{\ln{2}}$ in the
axisymmetric case, and  FWHM$_{s,x}$ = 2$\sigma_{s,x}\sqrt{\ln{2}}$
et FWHM$_{s,y}$ = 2$\sigma_{s,y}\sqrt{\ln{2}}$ in the elliptical case.
The beam profile is also Gaussian, with a beam size FWHM$_{obs}$ =
2$\sigma_{obs}\sqrt{\ln{2}}$. The true source size (FWHM$_{s}$) is
obtained by performing a deconvolution of the observed size using the
known beam profile.

The final line intensities is obtained by multiplying the
observed signal by the following scaling factor, which depends on the
source size, initial and final spatial resolution :

    \begin{center}
        \begin{equation}\label{fact-ronde-ge}
            fact=\frac{\theta_{mb,init}^{2}+\text{FWHM}_{s}^{2}}{\theta_{mb,fin}^{2}+\text{FWHM}_{s}^{2}}
        \end{equation}
    \end{center}

for the axisymmetric case, and :

    \begin{center}
        \begin{equation}\label{fact-ell-ge}
            fact=
            \sqrt{\frac{\theta_{mb,init}^{2}+\text{FWHM}_{s,x}^{2}}{\theta_{mb,fin}^{2}+\text{FWHM}_{s,x}^{2}}}\sqrt{\frac{\theta_{mb,init}^{2}+\text{FWHM}_{s,y}^{2}}{\theta_{mb,fin}^{2}+\text{FWHM}_{s,y}^{2}}}
        \end{equation}
    \end{center}

for elliptical sources.

\section{Tables of data}\label{tables}

\begin{longtable}[c]{|c|c|c|c|c|c|c|}
  \caption{Results from the gaussian fits derived from spectra of the Antennae
  (NGC 4038 et Overlap), Arp 220, Centaurus A, Henize 2-10, IC 10, IC 342,
  IRAS 10565, M 51, M 82, M 83, Mrk 231, NGC 253, NGC 891, NGC 1068, 3079,
  NGC 4736, NGC 6090 and NGC 6946. The observed lines ($^{12}$CO, $^{13}$CO
  and C) are indicated in the first column. For each source, the RA(1950)
  and DEC(1950) positions of the center (0",0") is listed in
  Table ~\ref{tab:prop}. The indication $^{a}$ corresponds to the
  spectra shown in this article (see Figs. ~\ref{fig:spec_ic10}
  to ~\ref{fig:spec_n6946_bras}).}\label{tab:obs1}\\
  \rowcolor{gris} Transition & Freq & beam & offset & $\int(T_{mb}dv)$ & Intensity & Flux \\
  \rowcolor{gris} &(GHz) & size(``) & position(``)& (Kkms$^{-1}$) & (Wm$^{-2}$sr$^{-1}$) & (Wm$^{-2}$) \\ \hline\endfirsthead
  \multicolumn{7}{|c|}{...continued...} \\ \hline
  \rowcolor{gris}Transition & Freq & beam & offset & $\int(T_{mb}dv)$ & Intensity & Flux \\
  \rowcolor{gris}&(GHz) & size(``) & position(``)& (Kkms$^{-1}$) & (Wm$^{-2}$sr$^{-1}$) & (Wm$^{-2}$) \\\hline\endhead
  \hline \multicolumn{7}{|c|}{...continued...} \\ \endfoot
  \hline \multicolumn{7}{r}{END} \\ \endlastfoot
  \hline
  \textbf{The Antennae}&&&&&&\\
  \textbf{NGC 4038}&&&&&&\\
  \hline
  CI($^{3}$P$_{1}$-$^{3}$P$_{0}$) & 492.162 &14.55 & (0,0)$^{a}$ & 13.7$\pm$2.1 & 1.7$\times10^{-9}$ & 9.4$\times10^{-18}$  \\
  \hline
  $^{12}$CO(2-1) & 230.538 & 30.50 & (0,0)$^{a}$ & 35.6$\pm$1.3 & 4.5$\times10^{-10}$ & 1.1$\times10^{-17}$  \\
  $^{12}$CO(3-2) & 345.796 & 21.90 & (-20,0)& 15.9$\pm$1.8 & 6.7$\times10^{-10}$ & 8.6$\times10^{-18}$  \\
  && 21.90 & (-10,0) & 29.5$\pm$1.4 & 1.2$\times10^{-9}$ & 1.6$\times10^{-17}$ \\
  && 21.90 & (0,0)$^{a}$ & 37.2$\pm$1.7 & 1.6$\times10^{-9}$ & 2.0$\times10^{-17}$ \\
  && 21.90 & (0,10) & 23.4$\pm$1.3 & 9.9$\times10^{-10}$ & 1.3$\times10^{-17}$ \\
\hline
\end{longtable}

\begin{longtable}[c]{|c|c|c|c|c|c|c|c|}
  \caption{Results of the gaussian fits derived from spectra of the Antennae
  (NGC 4038 et Overlap), Henize 2-10, IC 10, IC 342, M 83,
  NGC 253 and NGC 6946. See caption of Table ~\ref{tab:obs1}. Data
  listed in this table are either from literature or from our sample
  of observations.}\label{tab:obs2}\\
  \rowcolor{gris} Transition & Freq & beam & offset & $\int(T_{mb}dv)$ & Intensity & Flux & Ref. \\
  \rowcolor{gris} &(GHz) & size(``) & position(``)& (Kkms$^{-1}$) & (Wm$^{-2}$sr$^{-1}$) & (Wm$^{-2}$) & $^{a}$\\ \hline\endfirsthead
  \multicolumn{8}{|c|}{...continued...} \\ \hline
  \rowcolor{gris}Transition & Freq & beam & offset & $\int(T_{mb}dv)$ & Intensity & Flux & Ref. \\
  \rowcolor{gris}&(GHz) & size(``) & position(``)& (Kkms$^{-1}$) & (Wm$^{-2}$sr$^{-1}$) & (Wm$^{-2}$) & $^{a}$\\\hline\endhead
  \hline \multicolumn{8}{|c|}{...continued...} \\ \endfoot
  \hline \multicolumn{8}{r}{END} \\ \endlastfoot
  \hline
  \textbf{Antennae}&&&&&&&\\
  \textbf{NGC4038}&&&&&&&\\
  \hline
  CI($^{3}$P$_{1}$-$^{3}$P$_{0}$) & 492.162 & 14.55 & (0,0) & 13.7$\pm$2.1 & 1.7$\times10^{-9}$ & 9.4$\times10^{-18}$ & 1\\
  & & 21.90 & (0,0) & 7.6$\pm$1.2 & 9.2$\times10^{-10}$ & 1.2$\times10^{-17}$ &1*\\
  \hline
  $^{12}$CO(1-0) & 115.271 & 43.00 & (0,0) & 21.0$\pm$2.0 & 3.3$\times10^{-11}$ & 1.6$\times10^{-18}$ &2\\
  && 6.75 & (0,0) &415.0$\pm$83.0 & 6.5$\times10^{-10}$ & 7.9$\times10^{-19}$ &3\\
  & & 21.90 & (0,0) & 112.1$\pm$22.4 & 1.8$\times10^{-10}$ & 2.2$\times10^{-18}$ & 3\\
  & & 55.00 & (0,0) & 19.5$\pm$3.9 & 3.1$\times10^{-11}$ & 2.5$\times10^{-18}$ & 6\\
\hline
  \hline
\end{longtable}
\vspace{-0.5cm} \hspace*{-0.4cm}
\begin{minipage}{17cm}
\textbf{$^{a}$ References: \underline{Pour NGC 4038} : 1:
\citet{Geri00} but spectra have been analyzed again; 2:
\citet{Aalt95}; 3: \citet{Stan90}; 4: \citet{Zhu03}; 5: Our work; 6:
\citet{Gao01}; 7: \citet{Glen01};
*: used for constraining the models and for computing the C and CO cooling rates.\\
\underline{Pour Overlap} : 1: \citet{Geri00} but spectra have been
analyzed again; 2: \citet{Stan90}; 3: \citet{Aalt95}; 4:
\citet{Zhu03}; 5: Our work; 6: \citet{Gao01};
7: \citet{Glen01}; *: used for constraining the models and for computing the C and CO cooling rates..\\
\underline{Pour HENIZE 2-10} : 1: \citet{Baas94}; 2: \citet{Kobu95};
3: \citet{Meie01}; 4: \citet{Geri00} but spectra have been analyzed
again; 5: \citet{Baye04} ;
*: used for constraining the models and for computing the C and CO
cooling rates (see \citet{Baye04}).\\
\underline{Pour IC 10} : 1: \citet{Geri00} but spectra have been
analyzed again; 2: \citet{Bola00}; 3: Our work; 4: PhD Thesis of
Becker (1990); 5: \citet{Peti98a}; 6:
\citet{Glen01}; *: used for constraining the models and for computing the C and CO cooling rates..\\
\underline{Pour IC 342} : 1: \citet{Isra03}; 2: Our work; 3:
\citet{Ecka90c}; *: used for constraining the models and for computing the C and CO cooling rates..\\
\underline{Pour M 83} : 1: \citet{Peti98b}; 2: \citet{Isra01}; 3:
\citet{Lund04}; 4: \citet{Hand90}; 5: Our work; *: used for
constraining the models and for computing the C and CO cooling rates.\\
\underline{Pour NGC 253} : 1: \citet{Harr91}; 2: \citet{Wall91}; 3:
\citet{Gues93}; 4: \citet{Harr95}; 5: \citet{Isra95}; 6:
\citet{Maue96}; 7: \citet{Harr99}; 8: \citet{Dumk01}; 9:
\citet{Isra02}; 10: \citet{Brad03}; 11: \citet{Baye04}; *: used for
constraining the models and for computing the C and CO cooling rates (see \citet{Baye04}).\\
\underline{Pour NGC 6946} : 1: \citet{Isra01}; 2: \citet{Geri00} but
spectra have been analyzed again; 3: \citet{Wals02}; 4:
\citet{Ishi90}; 5: \citet{Weli88}; 6: \citet{Sofu88}; 7:
\citet{Caso90}; 8: Our work; 9: \citet{Clau91}; 10: \citet{Wall93};
11: \citet{Maue99}; 12: \citet{Niet99}; *: used for constraining the
models and for computing the C and CO cooling rates.\\}
\end{minipage}

\end{document}